\documentclass[preprint2]{aastex}

\begin{document}

\title{ Numerical Methods for the Simulation of Dynamical Mass Transfer in Binaries}
\author{Patrick M. Motl}
\affil{University of Missouri - Columbia}
\affil{Department of Physics and Astronomy, University of Missouri - Columbia, Columbia, MO 65211}
\email{motl@hades.physics.missouri.edu}
\author{Joel E. Tohline}
\email{tohline@rouge.phys.lsu.edu}
\and
\author{Juhan Frank}
\email{frank@rouge.phys.lsu.edu}
\affil{Louisiana State University}
\affil{Department of Physics and Astronomy, Louisiana State University, Baton Rouge, LA 70803}

\begin{abstract}
We describe computational tools that have been developed to simulate
dynamical mass transfer in semi-detached, polytropic binaries that are
initially executing synchronous rotation upon circular orbits.  Initial
equilibrium models are generated with a self-consistent field algorithm;
models are then evolved in time with a parallel, explicit, Eulerian
hydrodynamics code with no assumptions made about the symmetry of the
system.  Poisson's equation is solved along with the equations of ideal
fluid mechanics to allow us to treat the nonlinear tidal distortion of
the components in a fully self-consistent manner.  We present results
from several standard numerical experiments that have been conducted
to assess the general viability and validity of our tools, and from
benchmark simulations that follow the evolution of two detached systems
through five full orbits (up to approximately 90 stellar dynamical times).
These benchmark runs allow us to gauge the level of quantitative accuracy
with which simulations of semi-detached systems can be performed using
presently available computing resources.  We find that we should be able
to resolve mass transfer at levels $\dot{M} / {M} > \mathrm{few} \times  10^{-5}$ per
orbit through approximately 20 orbits with each orbit taking about
30 hours of computing time on parallel computing platforms.
\end{abstract}

\keywords{binaries: close, hydrodynamics, methods: numerical }

\section{Introduction}

Over half of all stars in the sky are actually multiple star
systems and, of the binaries, about half again are close enough to
one another for mass to be exchanged between the components at
some point in their evolution \citep{trimble83}.    There is a
subset of these close binary systems in which periodic or
aperiodic variations in luminosity and spectral features can be
explained by on-going mass-transfer events and instabilities in
the accretion flow. For example, long term stable mass transfer in
which the accretor is either a white dwarf, a neutron star, or a
black hole is widely recognized as the mechanism powering
cataclysmic variables \citep{warner95} and X-ray sources
\citep{lewin95}. In Algol--type systems an evolved star transfers
mass via Roche lobe overflow to a near main--sequence accretor
\citep{batten89,vesper01}. Each of these systems evolves on a
(secular) time scale that is long compared to the orbital period
of the system, with the mass-transfer rate determined by angular
momentum losses from the binary, and thermal relaxation and
nuclear evolution of the donor star.  In these systems, the
fraction of the donor's mass that is transferred during one orbit
is typically $\sim 10^{-12}$ to $10^{-9}$, many orders of
magnitude less than what current numerical 3-D hydrocodes can
resolve.  All of the above systems must have descended from
binaries in which the accretor of today was initially the more
massive component who evolved first off the main--sequence.  The
mass transfer in these progenitor systems was in many instances
dynamically unstable, yielding mass--transfer rates many orders of
magnitude above the currently observed rates and leading in some
cases to a common envelope phase (see e.g. Warner 1995; Verbunt \&
van den Heuvel 1995; Nelson \& Eggleton 2001)

In addition, there is a wide class of binary star systems
not presently undergoing mass-transfer for which the astrophysical
scenarios that have been proposed to explain their origin or
ultimate fate involve dynamical or thermal mass-transfer in a
close binary, sometimes leading to a common envelope phase of
evolution.  Examples of such systems are
millisecond pulsars \citep{bhattacharya95}, some central stars
of planetary nebulae \citep{iben93}, double degenerate
white dwarf binaries, perhaps leading to supernovae of type
Ia through a merger \citep{iben84}, or subdwarf sd0,
sdB stars \citep{iben90}, and double neutron star binaries,
perhaps yielding $\gamma$-ray bursts in a fireball when the
neutron stars coalesce \citep{paczynski86,ruffert97,meszaros01}.
The evolutionary scenarios that are drawn upon to explain the
existence of some of these systems call for events in which
$10\%$ or more of the donor's mass can be transferred during a single
orbit.

If we are to fully understand these rich classes of astrophysically
interesting systems --- their origin, present evolutionary state,
and ultimate fate --- it seems clear that we will have to develop
numerical algorithms that can accurately simulate mass-transfer
events in binary systems under a wide range of
physical conditions (for example, systems having a wide range of
total masses, mass ratios, and ages) over both short and long
evolutionary time scales.   The astrophysics community as a whole
is far from achieving this ultimate goal, but progress is being made
as various groups are methodically tackling small pieces of this very
large and imposing problem.  Examples of recent progress in the numerical
simulation of interacting binaries include two-dimensional simulation
of mass transfer in Algol \citep{blondin95}, three-dimensional
evolutions of Roche Lobe overflow in LMC X-4 \citep{boroson01}
and the accretion stream in $\beta$ Lyrae \citep{bisikalo00},
simulations of the common envelope phase and merger of a
point-mass white dwarf with a red giant star \citep{sandquist98},
neutron star binary and black hole-neutron star binary mergers in
the context of $\gamma$-ray bursts \citep{janka99} and the dispersal of
the secondary star's material in type Ia supernovae \citep{marieta00}.

Building on our experience simulating the nonlinear development of
dynamical instabilities in self-gravitating systems --- such as
protostellar gas clouds \citep{cazes00}, stellar cores
\citep{new00}, and young neutron stars \citep{lindblom01}
 --- and on our experience in studying mass-transfer in close binaries
through analytical and semi-analytical techniques
\citep{king97,frank01}
we are developing a hydrodynamical tool to study
relatively {\it rapid} phases of mass-transfer in binary systems.
Our immediate aim is to be able to follow the dynamical redistribution
of material through $\sim 10 - 30$ orbits after the onset of a
mass-transfer instability, in binary systems having a wide range of initial mass
ratios with either star initially selected to be in contact with
its Roche lobe and thereby become the ``donor''.
Our simulation tool treats both stars as self-gravitating fluids;
they are embedded in the computational grid in such a way that their
internal structures are both fully resolved; and the system as a
whole is evolved forward in time through an explicit integration of
the standard fluid equations, coupled with the Poisson equation so
that the Newtonian gravitational field changes along with the mass
distribution in a fully self-consistent way.   Initially we will
examine structures that can be well-represented by relatively simple
barotropic (and adiabatic) equations of state, but this constraint
can easily be lifted in the future.   We will be restricted to studies
of relatively rapid phases of mass-transfer because we are integrating
the equations of motion forward in time via an explicit integration scheme.

While this simulation tool will not permit us to model stable flows
with low mass-transfer rates --- such as the observed flows in
CVs and X-ray binaries --- it should be capable of a wide range of
astrophysical applications including:  A determination of the
conditions required to become unstable toward dynamical mass-transfer
in all kinds of close binaries with normal and degenerate components;
the ability to follow dynamical phases of mass transfer through to
completion, which may mean a return to stability at a new system
mass ratio, the formation of a massive disk or a common envelope with
or without rapid mass loss from the system,
or a merger of the two objects into one;
and an investigation of the steady-state structure of secularly
stable binaries.  Through such investigations we will be able to place
on much firmer footing a variety of theoretical scenarios (as alluded
to above) that have been proposed to explain the evolution and
fate of close binary systems.

With the commissioning of gravitational wave interferometers such as
TAMA, LIGO, and VIRGO, there has been a growing interest in understanding
the detailed behavior of, especially, neutron star inspirals and mergers.
As has been reviewed by Swesty, Wang and Calder (2000; hereafter SWC),
a number of different
groups have developed hydrodynamical codes to simulate the late stages
of inspiral and merger of such compact objects.  Indeed, as has been
described by \citet{new97} and reviewed by SWC,
an earlier version of our own simulation tool has been
used to study the dynamical merger of equal-mass systems in which the
stellar components were modeled with polytropic, white dwarf, and neutron
star equations of state.  Generally speaking, however, the last phase
of a neutron star inspiral can be modeled with a hydrodynamical code
that is less sensitive to initial conditions and more tolerant of errors
in the algorithm that integrates the fluid equations forward in time
than a hydrodynamical code that is designed to study more generic
mass-transfer events in close binary systems.  This is because general
relativistic effects will necessarily drive a binary neutron star system
to smaller separation, guaranteeing that the system will merge; and,
even in the absence of relativistic effects, it appears as though a tidal
instability that disrupts one or both stars will be encountered before
either fills its Roche lobe and encounters a classic
mass-transfer instability \citep{lai94}.

Building on the work of \citet{new97}, we now have a simulation
tool that can hydrodynamically follow the orbital evolution of
binary stars with high precision.  In developing this tool we have
made a number of improvements to the hydrodynamics algorithm that
was used in this earlier work only to study the tidal merger
problem.   We also have implemented a self-consistent-field
algorithm that can construct very accurate initial equilibrium
models of unequal-mass binaries in circular orbits, have paid
special attention to the manner in which initial models are
introduced into the hydrodynamics code, and have taken full
advantage of continuing improvements in high-performance
computers.  With this tool in hand, we should be able to
accurately model the evolution of a much broader class of close
binary systems, specifically, systems in which the components
initially have unequal mass and/or radii and in which a
mass-transfer instability rather than a tidal instability sets in.
With the inclusion of appropriate relativistic corrections, this
simulation tool should in principle also be able to simulate the
merger of equal or unequal mass neutron star binaries, but our
intention is not to focus so narrowly on this particular class of
systems.

In \S 2 of this paper, we collect results from theoretical investigations
of the linear stability of mass transfer in close binaries and
discuss the approximations that have been required to arrive at
these results.  In \S 3 we present the self-consistent field method
we use for the construction of initial, equilibrium models.  We
then describe our implementation of a parallel hydrodynamics code
for the solution of the ideal fluid equations and Poisson's equation
for an isolated mass distribution in \S 4.  In \S 5 we compare results
from the hydrodynamics code with known solutions for a set of test
problems and in \S 6 we present the results from the evolution of two
benchmark detached binaries.  These simulations demonstrate our
ability to faithfully represent the forces acting on the fluid and allow
us to estimate the mass transfer rate we will be able to resolve and
the computational expense required to evolve a given system through
an interesting number of orbits.  We conclude in \S 7 by summarizing
the limits we have been able to attain at practical simulation
resolutions and discuss the future application of the tool set to
systems of interest.

\section{Theoretical Description of Dynamical Mass Transfer}
\label{sec:stability}

In this section we will argue that Roche lobe overflow in a binary system approximated
by two polytropic components can result in mass transfer on a dynamical time
scale for a certain range of polytropic indices.  A spherical polytrope with uniform
entropy in mechanical equilibrium obeys the following mass radius relation
\citep[c.f.,][]{chandrasekhar39},
\begin{equation}
   \label{eq:poly_m_r}
   R \propto M^{\frac{1 - n}{3 - n}},
\end{equation}
which implies,
\begin{equation}
   \label{eq:xi_poly}
   \frac{\dot{R}}{R} = \left( \frac{1 - n}{3 - n} \right) \frac{\dot{M}}{M} \equiv
                       \xi_{\mathrm{S}} \frac{\dot{M}}{M}.
\end{equation}
Hence, the body will expand upon mass loss for polytropic indices
satisfying $1 < n < 3$.

Consider Paczy\'{n}ski's (1971) approximation for the effective Roche lobe radius
$R^{\mathrm{RL}}_{2}$
of a  donor star, taken to be the secondary, with mass $M_{2}$ in a point mass binary of total mass $M$ and
separation $a$,
\begin{equation}
   \label{eq:poly_log_r_m}
   \frac{R^{\mathrm{RL}}_{2}}{a} = 0.462
       \left( \frac{M_{2}}{M} \right)^{\frac{1}{3}}.
\end{equation}
From this, one obtains the following relation for the logarithmic change in the
donor's Roche lobe radius,
\begin{equation}
   \frac{\dot{R}^{\mathrm{RL}}_{2}}{R^{\mathrm{RL}}_{2}} =
        \frac{\dot{a}}{a} + \frac{1}{3}
        \frac{\dot{M}_{2}}{M_{2}} -
    \frac{1}{3} \frac{\dot{M}}{M}.
\end{equation}
Upon eliminating the separation in favor of the system's orbital angular
momentum $J$, one arrives at
\begin{equation}
   \frac{\dot{R}^{\mathrm{RL}}_{2}}{R^{\mathrm{RL}}_{2}}
       = 2 \frac{\dot{J}}{J} - \frac{5}{3}
        \frac{\dot{M}_{2}}{M_{2}} +
    \frac{2}{3} \frac{\dot{M}}{M}
    - 2 \frac{\dot{M}_{1}}{M_{1}},
\end{equation}
where $M_{1}$ is the mass of the accreting star, taken to be the primary.
If we further assume that the mass transfer is conservative with respect to the
total mass and orbital angular momentum we deduce that,
\begin{equation}
   \label{eq:log_roche_r}
   \frac{\dot{R}^{\mathrm{RL}}_{2}}{R^{\mathrm{RL}}_{2}} =
        \left( 2 \frac{M_{2}}{M_{1}} -
    \frac{5}{3} \right)
        \frac{\dot{M}_{2}}{M_{2}} \equiv
    \xi_{\mathrm{R}} \frac{\dot{M}_{2}}{M_{2}}.
\end{equation}
Comparing equation (\ref{eq:xi_poly}) with equation (\ref{eq:log_roche_r}),
the condition for stable mass transfer, $\dot{R}_{2} \leq
\dot{R}^{\mathrm{RL}}_{2}$ can be expressed as,
\begin{equation}
   \xi_{\mathrm{S}} - \xi_{\mathrm{R}} > 0,
\end{equation}
which, for a given polytropic index,  implies a stable mass ratio
\begin{equation}
   q \equiv \frac{M_{2}}{M_{1}} =
   q_{\mathrm{stable}} \equiv \frac{9 - 4n}{3 \left( 3 - n \right)}.
\end{equation}
For a polytropic binary with $n = 3 / 2$ and mass ratio
$q > q_{\mathrm{stable}} = 2 / 3$,
mass transfer must occur on a dynamical time scale
as the donor will readjust its structure within a few sound crossing times to its new mass.
Note that if the donor is initially the less massive star (i.e.\, $q < 1$), the binary
separation is expected to steadily increase during the mass transfer event.
But, if the donor is initially the more massive component (i.e.\, $q > 1$), conservation
of orbital angular momentum implies that the separation must decrease and that the donor's
Roche lobe radius will contract thus increasing the degree of overflow.
The resulting mass transfer rate is expected in this case to be quite substantial.

The dependence of the mass transfer rate on the degree of over-contact can be
estimated from the product of the volume swept out by the flow near the
inner Lagrange point, $L_{\mathrm{1}}$,
in unit time and the local value of the density.  The cross section of the flow
near $L_{\mathrm{1}}$ will scale as the square of the local sound speed, and the
flow velocity is approximately equal to the sound speed.  The volume of material
transferred in unit time then scales as the cube of the sound speed.  The density
near the edge of a spherical polytrope of index $n$, radius $R_{2}$, mass
$M_{2}$, and polytropic constant $\kappa$ can be found by integrating the
equation of hydrostatic equilibrium to obtain,
\begin{equation}
   \rho \left( r \right) \approx \left[ \frac{G M_{2}}{\kappa \left( n + 1 \right)}
                                  \frac{\left( R_{2} - r \right)}{ R^{2}_{2}} \right]^{n}.
\end{equation}
If we change variables to $\Delta R_{2} \equiv R_{2} - r$, the width of a spherical shell near the
edge of the star, we obtain
\begin{equation}
   \rho \left( \Delta R_{2} \right) \propto \left( \Delta R_{2} \right)^{n}.
\end{equation}
The sound speed, $c$ in turn obeys,
\begin{equation}
   c \propto \rho^{\frac{1}{2n}},
\end{equation}
so that
\begin{equation}
   \frac{d M_{2}}{d t} \propto \rho^{1 + \frac{3}{2 n}} \propto
          \left( \Delta R_{2} \right)^{n + \frac{3}{2}}.
\end{equation}
Taking the radius, $r$, to be the effective Roche lobe radius of
the donor, $\Delta R_{2}$ is the degree of over-contact.  The mass
transfer rate is then expected to scale with the degree of
over-contact as
\begin{equation}
  \label{eq:mdot}
   \frac{d M_{2}}{d t} \propto \frac{M_{2}}{P}
    \left( \frac{\Delta R_{2}}{R_{2}} \right)^{n + \frac{3}{2}},
\end{equation}
where $P$ is the orbital period.  This agrees with the calculation
of Jedrzjec as presented in \citet{paczynski72}.  For a polytropic
index of $n = 3 / 2$, eq.\ (\ref{eq:mdot}) indicates that
the mass transfer rate will scale as the cube of the degree of
over-contact. While the actual mass transfer rate observed in a
fully self-consistent, three-dimensional evolution may differ
substantially from the estimate given in (\ref{eq:mdot}), it
nevertheless indicates that for unstable binaries mass-transfer
events will evolve on a dynamical time scale once the donor
reaches contact with its Roche lobe.

All the results presented in this section have relied on a great
many simplifying assumptions including the disregard of internal
angular momentum (spin) in each star, the use of the Roche model,
neglecting the intrinsically nonspherical geometry of the
components, and assuming that the mass transfer event is, in fact,
conservative.  To proceed beyond this point one must deal with
extended distributions for the density and velocity in some
approximation.  We would argue further that it is advantageous to
use a potential derived from the matter distribution in a
self-consistent manner. With these additional complications, the
task is well beyond the regime of analytical mechanics, but is
tractable if we employ three-dimensional computational fluid
dynamical techniques.  To investigate short time-scale mass
transfer events numerically, we have developed a set of tools for
both constructing equilibrium polytropic binaries and a
hydrodynamics code to evolve systems of interest in time.  These
tools are described below.

\section{Construction of Equilibrium Models}
\label{sec:scf}

The iterative method that we have used to generate equilibrium,
polytropic binaries is very closely related to the self-consistent
field (SCF) technique developed by Hachisu (1986; see also Hachisu,
Eriguchi \& Nomoto  1986). This technique previously has been used to construct
initial models of {\it equal}-mass binary systems for dynamical
studies of the binary merger problem (see New \& Tohline 1997 and
SWC). Here we employ a more generalized version of the technique
to construct {\it unequal}-mass binaries. In the following
discussion we use $\mathbf{r}$ to refer to an arbitrary point in
space.  The vector $\mathbf{R}$ is the cylindrical radius vector
which can be expressed as $\mathbf{R} = x \, \hat{\mathbf{x}} + y
\, \hat{\mathbf{y}}$.  The axis of rotation is always taken to be
parallel to, but not necessarily coincident with, the $z$-axis.

The numerical results presented here are in a system of units
where the gravitational constant, the radial extent of our
numerical grid and the maximum density of one binary component are
all taken to be unity.  As we are treating polytropic models
exclusively in the present work, these models can be scaled to
represent different physical systems by choosing a total system
mass or orbital separation for example. We would like to emphasize
that the polytropic models could represent binaries consisting of
neutron stars, white dwarfs or normal stellar components.

Assuming synchronous rotation so that the bodies are stationary in
a corotating reference frame, the equations of hydrostatic
equilibrium reduce to the following  single vector equation,
\begin{equation}
   \mathbf{\nabla} \left( H + \Phi - \frac{1}{2} \Omega^{2}
   \left| \mathbf{R} - \mathbf{R}_{\mathrm{com}} \right|^{2} \right) = 0.
   \label{eq:euler_equation_static}
\end{equation}
Here $\Phi$ is the gravitational potential, $\Omega$ is the
angular velocity of the reference frame in which the fluid is
stationary, H is the enthalpy, and $\mathbf{R}_{\mathrm{com}}$ is
the cylindrical radius vector of the system's center of mass so
that $\left| \mathbf{R} - \mathbf{R}_{\mathrm{com}} \right|$ is
each fluid element's distance from the axis of rotation. For a
polytrope of index $n$, $H$ is given to within an arbitrary
constant by
\begin{equation}
   H = \int \frac{dp}{\rho} = \left( n + 1 \right) \frac{p}{\rho} =
       \left( n + 1 \right) \kappa \rho^\frac{1}{n},
\end{equation}
where $p$ and $\rho$ are the pressure and density of the fluid,
respectively. Equation (\ref{eq:euler_equation_static}) in turn
implies that,
\begin{equation}
   H + \Phi - \frac{1}{2} \Omega^{2} \left| \mathbf{R} - \mathbf{R}_{\mathrm{com}}
   \right|^{2} = C,
   \label{eq:iteration}
\end{equation}
for some constant $C$, which in general will be different for each
binary component.  Hereafter we denote these two integration
constants as $C_{\mathrm{1}}$ and $C_{\mathrm{2}}$.

Using equation (\ref{eq:iteration}) one can construct an iterative
scheme as follows. An initial guess at the density field is
constructed. Poisson's equation is solved to obtain the
gravitational potential arising from the chosen mass distribution.
This is, by far, the most computationally intensive part of the
algorithm. For our work, we have chosen a cylindrical coordinate
grid and utilized subroutines from the FISHPACK Fortran subroutine
set for the solution of elliptic partial differential equations
\citep{schwarztrauber75, schwarztrauber00} with the boundary
potential being calculated via a spherical harmonic expansion of
the density field utilizing harmonic moments through $\ell = 10$.

\begin{figure}
   \plotone{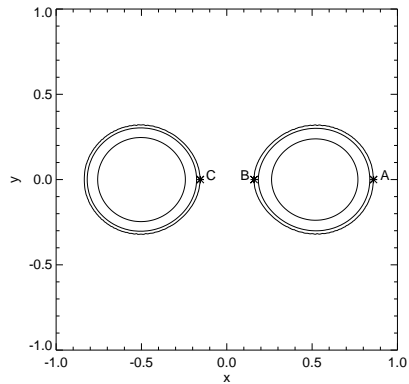}
   \figcaption[f1.eps]{Position of the three boundary points in the equatorial plane for a
   SCF binary model.  At the three boundary points (denoted by asterisks and labeled A, B, and C) the
   density is forced to vanish.  The contours represent density levels for the converged
   model.\label{fig:boundary_points}}
\end{figure}

With the gravitational potential in hand we can use algebraic
relations at three boundary points where the density field is
forced to vanish in order to set the integration constants
$C_{\mathrm{1}}$ and $C_{\mathrm{2}}$, and the value of the
angular velocity $\Omega$.  The boundary points all lie along the
line of centers.  They correspond to the inner and outer boundary
points for one star, and the inner boundary point for the
companion as illustrated in Fig.\  \ref{fig:boundary_points}. The
values of the gravitational potential at the three boundary
points, $\mathbf{r}_{\mathrm{A}}$, $\mathbf{r}_{\mathrm{B}}$ and
$\mathbf{r}_{\mathrm{C}}$ are used to solve for the set of
constants $\Omega$, $C_{1}$ and $C_{2}$ as follows:
\begin{equation}
    \Omega^{2} = \frac{\Phi \left( \mathbf{r}_{\mathrm{A}} \right) -
                    \Phi \left( \mathbf{r}_{\mathrm{B}} \right)}
           { \frac{1}{2} \left( \left| \mathbf{R}_{\mathrm{A}} -
                                        \mathbf{R}_{\mathrm{com}} \right| ^{2} -
                                        \left| \mathbf{R}_{\mathrm{B}} -
                           \mathbf{R}_{\mathrm{com}} \right| ^{2} \right) },
\end{equation}
\begin{equation}
    C_{1} = \Phi \left( \mathbf{r}_{\mathrm{B}} \right) - \frac{1}{2} \Omega^{2}
            \left| \mathbf{R}_{\mathrm{B}} - \mathbf{R}_{\mathrm{com}} \right| ^{2},
\end{equation}
\begin{equation}
    C_{2} = \Phi \left( \mathbf{r}_{\mathrm{C}} \right) - \frac{1}{2} \Omega^{2}
            \left| \mathbf{R}_{\mathrm{C}} - \mathbf{R}_{\mathrm{com}} \right| ^{2}.
\end{equation}
Equation (\ref{eq:iteration}) can then be used to construct the enthalpy
throughout the computational domain and, from it,
an improved density distribution can be constructed using the relation,
\begin{equation}
   \rho = \left\{
      \begin{array}{lr}
          \rho_{\mathrm{max}, i} \left( \frac{H}{H_{\mathrm{max}, i}} \right)^{n} & H > 0 \\
           0 &  \mathrm{otherwise} \\
      \end{array}
   \right.,
\end{equation}
where $i = 1,2$ labels the two stellar components.  As
\citet{hachisu86} has explained, it is best to hold the values of
$\rho_{\mathrm{max}, 1}$ and $\rho_{\mathrm{max}, 2}$ fixed
throughout the iteration cycles.

The iteration cycle is then repeated using the improved density
distribution until the relative change from iteration to iteration
in $C_{\mathrm{1}}$, $C_{\mathrm{2}}$, $\Omega$, and
$H_{\mathrm{max}, i}$ are all smaller than some prescribed
convergence criterion, $\delta$.  For a grid resolution of 128
radial points by 128 vertical points by 256 points in azimuth, we
typically use a tolerance $\delta = 1 \times 10^{-4}$.

Unfortunately the self-consistent field method does not allow one
to specify physically meaningful parameters such as the binary
mass ratio or separation \textit{a priori}.  Instead, as already
described, it is best to specify the three boundary points and the
maximum density for each body.  Nevertheless, the method described
above remains, to our knowledge, the most effective means of
generating fully self-consistent models of synchronously rotating,
equilibrium binary systems with unequal masses and/or radii.

We gauge the quality of a converged solution by the degree to
which it satisfies the scalar virial equation. Specifically, we
define the following dimensionless virial error
\begin{equation}
   \mathrm{VE} \equiv \frac{\left( 2 K + W + 3 \Pi \right)}{\left| W \right|},
   \label{eq:virial-error}
\end{equation}
where the terms appearing in equation (\ref{eq:virial-error}) are defined by the
following integral quantities:
\begin{equation}
   \label{eq:t}
   K \equiv \frac{1}{2} \int \rho \, \mathbf{v} \cdot \mathbf{v} \, dV,
\end{equation}
\begin{equation}
   \label{eq:w}
   W \equiv \int \rho \, \Phi \,  dV,
\end{equation}
\begin{equation}
   \label{eq:pi}
   \Pi \equiv \int p \, dV,
\end{equation}
where $\mathbf{v}$ is the velocity field as measured in the
inertial frame of reference.  As applied in the SCF technique, the
velocity is entirely due to the rotation of the frame, that is
\begin{equation}
\mathbf{v} = \mathbf{\Omega} \times
             \left( \mathbf{R} - \mathbf{R}_{com} \right).
\end{equation}

\begin{figure}
    \plotone{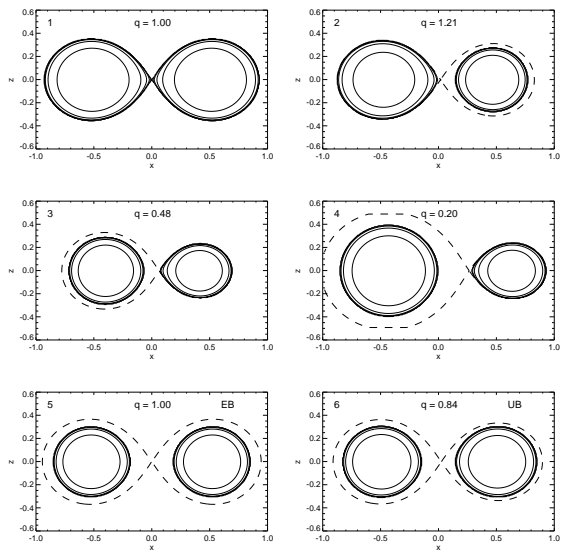}
    \figcaption[f2.eps]{Slice through the meridional plane
    for six example SCF binaries.  Detailed parameters for these models are provided in Table
    \ref{tab:scf_models}.  The solid contours are in the logarithm of the normalized density
    at levels of $10^{-5}$, $10^{-4}$, $10^{-3}$, $10^{-2}$, and $10^{-1}$.  The dashed curve
    traces the critical surface for the self-consistent Roche potential.  The density of the
    stellar components falls rapidly near the surface and on the scale used for these figures,
    the lowest two contours of the density coincide with one another.\label{fig:scf_side}}
\end{figure}

\begin{figure}
   \plotone{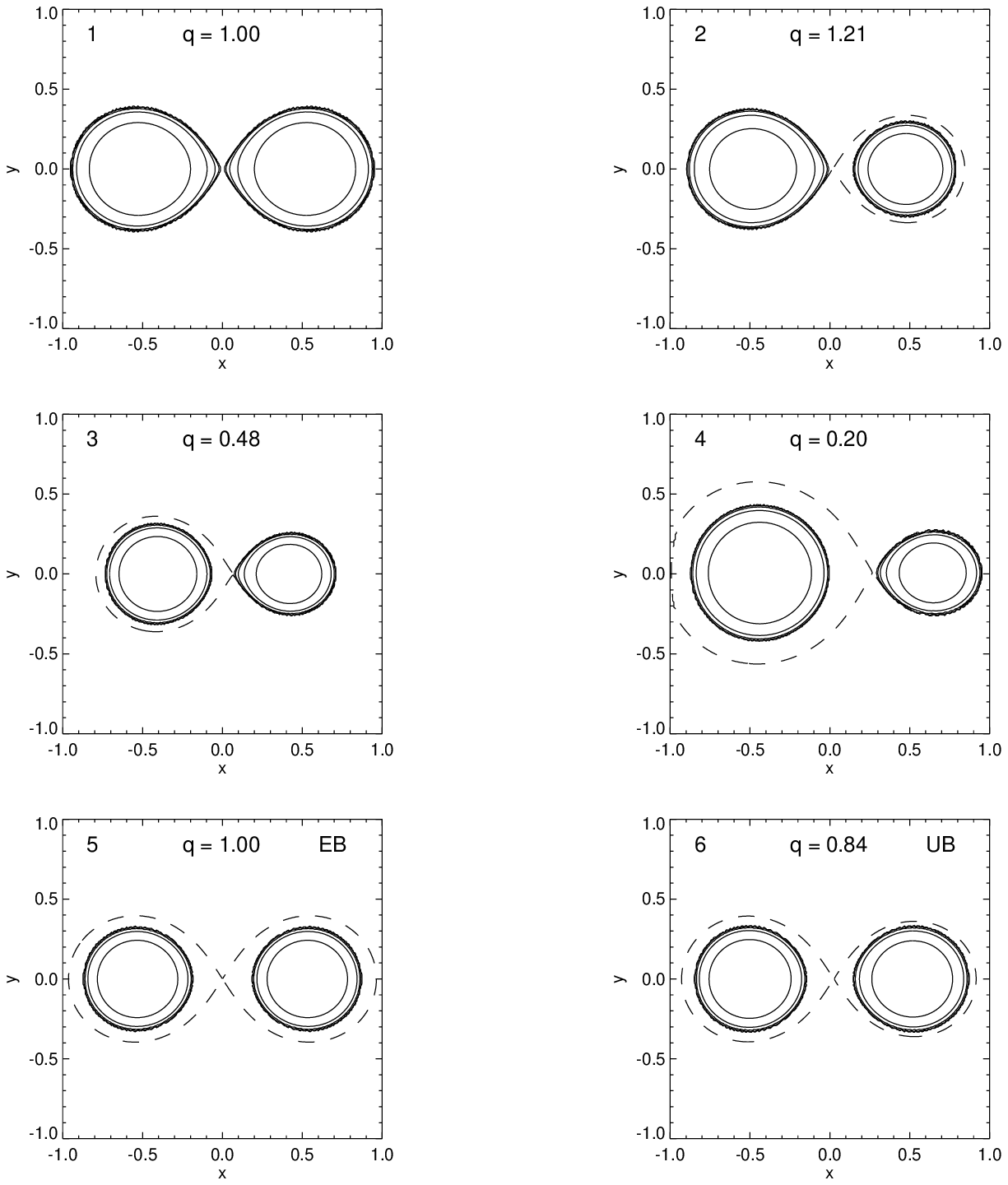}
   \figcaption[f3.eps]{Slice through the equatorial plane for the same six binaries shown in
   Fig.\  \ref{fig:scf_side}.  The solid contours are in the logarithm of the normalized density
   at levels of $10^{-5}$, $10^{-4}$, $10^{-3}$, $10^{-2}$ and $10^{-1}$.  The dashed curve
   traces the critical surface for the self-consistent Roche potential. \label{fig:scf_top}}
\end{figure}

In Fig.\  \ref{fig:scf_side} we plot density contours in the
meridional plane for one contact binary system, three
semi-detached systems, and two detached systems that we have
constructed using the SCF technique. The more massive component
always appears on the left of the plots. Figure  \ref{fig:scf_top}
shows contours in the equatorial plane for the same six systems.
The solid lines are at mass density levels of $10^{-5}$,
$10^{-4}$, $10^{-3}$, $10^{-2}$ and $10^{-1}$, where the density
has been normalized to the maximum density for each model, and the
dashed line follows the self-consistently determined critical
Roche surface for the system.  The binaries all have a polytropic
index $n = 3 / 2$; other key parameters for these models are
listed in Table \ref{tab:scf_models}.  Throughout this work we
take the secondary component (denoted by a `2') to be the
component closest to contact (closest to being the donor).
The values of $q = M_2/M_1$ shown in
Table \ref{tab:scf_models} (as well as in Figs.\
\ref{fig:scf_side} and \ref{fig:scf_top}) give the ratio of the
mass of the secondary to the primary; the stellar radii ($R_{1}$
and $R_{2}$) and Roche lobe radii ($R^{\mathrm{RL}}_{1}$ and
$R^{\mathrm{RL}}_{2}$) have all been normalized to the orbital
separation.  The stellar radii and Roche lobe radii are the radii
of spheres that have a volume equal to the star or critical Roche
surface, respectively. For Model 4, the Roche lobe of the primary
extends beyond the computational grid so the effective radius of
its Roche lobe is only a lower limit.  All six of these models
were constructed on a cylindrical grid of 128 radial and vertical
zones by 256 azimuthal zones.

\begin{deluxetable}{lrrrrrrrr}
   \tablenum{1}
   \tablewidth{0pt}
   \tablecolumns{9}
   \tablecaption{Initial Equilibrium Binary Models\label{tab:scf_models}}
   \tablehead{
                \colhead{Model} &
                \colhead{$q$} &
                \colhead{$\rho_{1}^{\mathrm{max}}$} &
                \colhead{$R_{1}$} &
                \colhead{$R^{\mathrm{RL}}_{1}$} &
                \colhead{$\rho_{2}^{\mathrm{max}}$} &
                \colhead{$R_{2}$} &
                \colhead{$R^{\mathrm{RL}}_{2}$} &
                \colhead{VE} }

   \startdata
         1      & 1.0000 & 1.00 & 0.3720 & 0.3723 & 1.00 & 0.3720 & 0.3723 & $1.5 \times 10^{-4}$\\
         2      & 1.2111 & 1.00 & 0.3056 & 0.3580 & 0.60 & 0.3893 & 0.3915 & $1.4 \times 10^{-4}$\\
         3      & 0.4801 & 1.20 & 0.3727 & 0.4401 & 1.00 & 0.3126 & 0.3129 & $3.4 \times 10^{-4}$\\
         4      & 0.1999 & 1.00 & 0.3817 & $>$ 0.5194 & 0.77 & 0.2476 & 0.2478 & $2.8 \times 10^{-4}$ \\
         5 (EB) & 1.0000 & 1.00 & 0.2984 & 0.3778 & 1.00 & 0.2984 & 0.3778 & $2.0 \times 10^{-4}$ \\
         6 (UB) & 0.8436 & 1.20 & 0.3180 & 0.3919 & 1.00 & 0.3200 & 0.3620 & $2.2 \times 10^{-4}$
    \enddata
\end{deluxetable}

Table \ref{tab:scf_convg} lists the resulting virial error for the
contact system (Model 1 from Table \ref{tab:scf_models})
constructed on grids of differing resolutions.  As the convergence
criterion, $\delta$, is decreased, the number of required
iterations increases.  For fixed resolution, the overall quality
of the solution does not significantly improve beyond some
limiting value of $\delta$, regardless of the number of iterations
taken.  As the resolution is increased, the virial error decreases
roughly in proportion to the square root of the number of grid
points.

\begin{deluxetable}{ccccc}
    \tablewidth{0pt}
    \tablenum{2}
    \tablecolumns{5}
    \tablecaption{Convergence for SCF Method\label{tab:scf_convg}}
    \tablehead{
        \colhead{R} & \colhead{z} & \colhead{$\phi$} & \colhead{$\delta$} & \colhead{VE} }
			       
        \startdata
          64  &  64 & 128 & $1.0 \times 10^{-3}$ & $1.0 \times 10^{-3}$ \\
              &     &     & $1.0 \times 10^{-4}$ & $6.0 \times 10^{-4}$ \\
              &     &     & $1.0 \times 10^{-5}$ & $5.5 \times 10^{-4}$ \\
              &     &     & $1.0 \times 10^{-6}$ & $5.5 \times 10^{-4}$ \\
          128 & 128 & 256 & $1.0 \times 10^{-3}$ & $6.9 \times 10^{-4}$ \\
              &     &     & $1.0 \times 10^{-4}$ & $2.0 \times 10^{-4}$ \\
              &     &     & $1.0 \times 10^{-5}$ & $1.5 \times 10^{-4}$ \\
              &     &     & $1.0 \times 10^{-6}$ & $1.4 \times 10^{-4}$ \\
          256 & 256 & 512 & $1.0 \times 10^{-3}$ & $6.3 \times 10^{-4}$ \\
              &     &     & $1.0 \times 10^{-4}$ & $1.0 \times 10^{-4}$ \\
              &     &     & $1.0 \times 10^{-5}$ & $5.2 \times 10^{-5}$ \\
              &     &     & $1.0 \times 10^{-6}$ & $4.7 \times 10^{-5}$ \\
              &     &     & $1.0 \times 10^{-7}$ & $4.7 \times 10^{-5}$
         \enddata
\end{deluxetable}

Due to the symmetry of these initial models about the equatorial
plane, we only calculate the models in the half space of $z \geq
0$.   Assuming the line of centers coincides with the $x$ axis,
the tidal distortion of each star also is symmetric about the $y =
0$ plane. Hence, further computational efficiency could be
obtained with this technique by limiting the computational grid to
only extend from 0 to $\pi$ in azimuth. To date, we have not
enforced this additional symmetry constraint, although in practice
the converged models display this symmetry.

The SCF method is insensitive to the functional form for the
initial guess of the density distribution.  For uniform spheres
and spherically symmetric Gaussian density distributions one can
obtain the same converged model to machine accuracy. We also note
that we have found that more rapid convergence for models with
soft equations of state (\textit{e.g.}, $n \geq 3 / 2$) can
be achieved by using an even mixture of the current and previous
potentials during the iteration.  For more rigid equations of
state, where there is more mass at the boundary points and hence a
greater coupling between the solution near the boundary points and
the global solution, such mixing is not necessary and the solution
converges rapidly.

\section{Hydrodynamics Implementation}

\subsection{Continuum Mechanics Formalism}
\label{sec-analytic}

We have developed an explicit, conservative, finite-volume,
Eulerian hydrodynamics code that is second-order accurate in both
time and space to evolve the equilibrium binaries.  The program is
similar to the ZEUS code developed by  \citet{stone92}. The
integration scheme is designed to evolve five primary variables
that are densities of conserved quantities: the mass density,
$\rho$, the angular momentum density, $A$, the radial momentum
density, $S$, the vertical momentum density, $T$, and an entropy
tracer, $\tau$. The entropy tracer,
\begin{equation}
  \tau \equiv \left( \epsilon \rho \right)^{\frac{1}{\gamma}},
\end{equation}
where $\epsilon$ is the internal energy per unit mass and $\gamma$ is
the selected ratio of specific heats of the gas.
It is related to the entropy of the fluid through the
relation,
\begin{equation}
   \label{eq:entropy}
   s = c_{\mathrm{p}} \ln{ \frac{\tau}{\rho} },
\end{equation}
where $c_{\mathrm{p}}$ is the specific heat at constant pressure.
Using the entropy tracer in lieu of the internal energy per unit
mass or the total energy density allows us to avoid the finite
difference representation of the divergence of the velocity field
that must otherwise be used to express the work done by pressure
on the fluid.

For the evolutions presented in this paper we have set
$\gamma = 1 + 1/n$.  We note, however, that by allowing the
compressible fluid system to evolve with an adiabatic exponent
that differs from this value, the stars will not be homentropic.
For example, by selecting an appropriate value of
$\gamma > 1 + 1/n$ we can effectively model stars that are
convectively stable and that obey a mass-radius relation quite
different from the normal polytropic one specified by
eq.\ (\ref{eq:xi_poly}), that is, different from
$\xi_{\mathrm{S}} = (1-n)/(3-n)$.  By doing this, we expect to
be able to closely approximate the mass-radius relationship of
main sequence stars.  A more detailed discussion of this idea
is beyond the scope of this paper.

The set of differential equations that we solve is based on
the conservation laws for these five conserved densities.
Mass conservation is governed by the continuity equation,
\begin{equation}
   \frac{\partial \rho}{\partial t} + \mathbf{\nabla} \cdot
        \left( \rho \, \mathbf{v} \right) = 0,
\end{equation}
where $\mathbf{v}$ is the velocity field.  The velocity vector is
expressed in terms of its components in a cylindrical coordinate
system as $\mathbf{v} = u \, \hat{\mathbf{e}}_{R} + v \,
\hat{\mathbf{e}}_{\phi} + w \, \hat{\mathbf{e}}_{z}$. The three
components of Euler's equation govern changes in the momentum
densities.  We express these equations in a frame of reference
rotating with a constant angular velocity, $\Omega$, as follows:
\begin{equation}
   \frac{\partial S}{\partial t} + \mathbf{\nabla} \cdot
        \left( S \mathbf{v} \right) =
   - \rho \frac{\partial \Phi^{\mathrm{eff}}}{\partial R}
   + \frac{A^{2}}{\rho R^{3}} + 2 \Omega \frac{A}{R},
   \label{eq:radial-euler-eq}
\end{equation}
\begin{equation}
   \frac{\partial T}{\partial t} + \mathbf{\nabla} \cdot
        \left( T \mathbf{v} \right) =
    - \rho \frac{\partial \Phi^{\mathrm{eff}}}{\partial z},
    \label{eq:vertical-euler-eq}
\end{equation}
\begin{equation}
   \frac{\partial A}{\partial t} + \mathbf{\nabla} \cdot
        \left( A \mathbf{v} \right) =
   - \rho \frac{\partial \Phi^{\mathrm{eff}}}{\partial \phi}
   - 2 \Omega S R,
   \label{eq:angular-euler-eq}
\end{equation}
where,
\begin{equation}
   \Phi^{\mathrm{eff}} \equiv H + \Phi - \frac{1}{2} \Omega^{2} R^{2}.
   \label{eq:effective-potential}
\end{equation}
The second and third terms appearing on the right-hand side of
eq.\ (\ref{eq:radial-euler-eq}) represent the curvature of
cylindrical coordinates and the radial component of the Coriolis
force, respectively.  Likewise, the last term appearing on the
right-hand side of eq.\  (\ref{eq:angular-euler-eq}) represents
the azimuthal component of the Coriolis force.

From the first law of thermodynamics we know that in the most
general case, the entropy tracer obeys the expression,
\begin{equation}
   \frac{\partial \tau}{\partial t} + \mathbf{\nabla} \cdot
        \left( \tau \mathbf{v} \right) = \frac{\tau}{c_{\mathrm{p}}} \frac{ds}{dt}.
\end{equation}
Here we will be considering only adiabatic flows, in which case
$d s / d t = 0$, so the entropy tracer obeys an advection
equation of precisely the same form as the continuity equation,
namely,
\begin{equation}
   \frac{\partial \tau}{\partial t} + \mathbf{\nabla} \cdot
        \left( \tau \mathbf{v} \right) = 0.
   \label{eq:entropy-tracer}
\end{equation}
Even though we are performing adiabatic evolutions we can not
simply use an adiabatic equation of state ($p = \kappa
\rho^{\gamma}$) and disregard the first law of thermodynamics
because the polytropic constant is, in general, different for each
binary component.

Finally, we solve Poisson's equation once every integration
timestep in order to calculate the force of gravity arising from
the instantaneous mass distribution,
\begin{equation}
   \nabla^{2} \Phi = 4 \pi G \rho;
\end{equation}
and we use the ideal gas law as the equation of state to close the
system of equations, namely,
\begin{equation}
   p = \left( \gamma - 1 \right) \tau^{\gamma} = \left( \gamma - 1 \right) \rho \epsilon.
\end{equation}

It may be argued that our treatment of the thermodynamics of the
system as the purely adiabatic flow of an ideal fluid is overly
simplified. However, we believe that the self-consistent treatment
of both binary components in the presence of the full nonlinear
tidal forces is sufficiently complex and novel to warrant the use
of a simple equation of state at the present time.  This will
allow us to establish the qualitative behavior of systems in this
limiting case before additional complications leading to
nonadiabatic heat transport are introduced into the simulations.

\subsection{Finite Volume Representation}

\begin{figure}
   \plotone{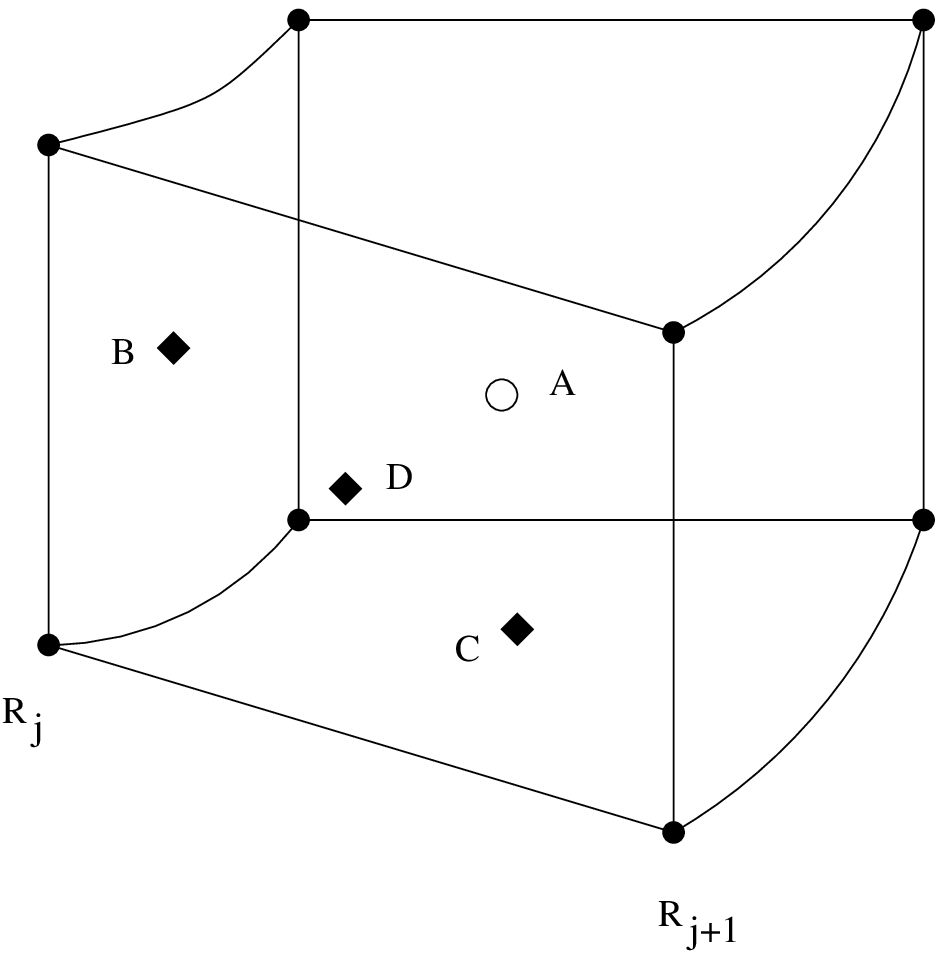}
   \figcaption[f4.eps]{Volume element for a cell-centered quantity (defined at the open circle labeled A)
   in our cylindrical coordinate mesh.  Radial, vertical and azimuthal face-centered quantities are defined 
   at points B, C, and D, respectively.  Table \ref{tab:variable_centering} lists all of the variables 
   used in the hydrodynamics code along with their centering according to this diagram.\label{fig:cell}}
\end{figure}

Before proceeding with the discussion of the hydrodynamics
algorithm that we have implemented to solve the equations
presented in \S \ref{sec-analytic} we first describe the
discretization that has been used to represent the exact partial
differential equations when they are expressed as approximate
algebraic relations between discrete points in the computational
grid.  As in the ZEUS code, all scalar variables and the diagonal
components of tensors are defined at cell centers.  The components
of vectors are defined at the corresponding faces of the cell.  A
volume element and the relative positions of the variables within
each cell is illustrated in Fig.\ \ref{fig:cell}.  The cell
extends from $R_{i}$ to $R_{i+1}$ in radius, from $z_{j}$ to
$z_{j+1}$ in the vertical coordinate, and from $\phi_{k}$ to
$\phi_{k+1}$ in the azimuthal coordinate.  We represent the
staggered variables in the computational mesh with a half-index
notation; the coordinates of the center of a grid cell are given
by $R_{i+1/2}$, $z_{j+1/2}$, $\phi_{k+1/2}$, for example. A
complete listing of the variables and their centering is given in
Table \ref{tab:variable_centering}.

\begin{deluxetable}{cll}
   \tablenum{3}
   \tablewidth{0pt}
   \tablecolumns{3}
   \tablecaption{Hydrodynamic Variables and their Centering\label{tab:variable_centering}}
   \tablehead{
      \colhead{Centering} & \colhead{Variable} & \colhead{Description} }

      \startdata
        A & $R_{i+1/2}$   & Cylindrical Radius Coordinate \\
          & $z_{j+1/2}$   & Vertical Coordinate \\
          & $\phi_{k+1/2}$   & Azimuthal Coordinate \\
          & $\rho_{i+1/2,j+1/2,k+1/2}$ & Mass Density \\
          & $\tau_{i+1/2,j+1/2,k+1/2}$ & Entropy Tracer \\
          & $p_{i+1/2,j+1/2,k+1/2}$    & Pressure \\
          & $H_{i+1/2,j+1/2,k+1/2}$    & Enthalpy \\
          & $\Phi_{i+1/2,j+1/2,k+1/2}$ & Gravitational Potential \\
          & $Q^{ll}_{i+1/2,j+1/2,k+1/2}$ & Diagonal Components of Artificial Viscosity \\
        B & $S_{i,j+1/2,k+1/2}$   & Radial Momentum Density \\
          & $u_{i,j+1/2,k+1/2}$   & Radial Velocity \\
        C & $T_{i+1/2,j,k+1/2}$   & Vertical Momentum Density \\
          & $w_{i+1/2,j,k+1/2}$   & Vertical Velocity \\
        D & $A_{i+1/2,j+1/2,k}$   & Angular Momentum Density \\
          & $v_{i+1/2,j+1/2,k}$   & Azimuthal Velocity
       \enddata
\end{deluxetable}

\subsection{Treatment of Advection Terms}

Through the method of operator splitting, one can construct a
numerical scheme that groups terms of the same physical character
together.  Again, following along the lines of the ZEUS code we
implement a splitting scheme that separates updates of the fluid
state due to Eulerian transport (advection) from updates due to
the source terms. In this section we describe our treatment of the
advection terms.

Given the density $\lambda$ of any conserved quantity $\Lambda$
that satisfies a generic conservation law of the form,
\begin{equation}
   \frac{\partial \lambda}{\partial t} + \mathbf{\nabla} \cdot \left( \lambda \, \mathbf{v} \right) = 0,
   \label{eq:advection-equation}
\end{equation}
we can replace the differential equation
(\ref{eq:advection-equation}) with an equivalent integral
equation,
\begin{equation}
   \frac{\partial}{\partial t} \int_{V} \lambda \, dV =
   - \int_{V} \mathbf{\nabla} \cdot \left( \lambda \, \mathbf{v} \right) dV =
   - \int_{S \left( V \right)} \lambda \, \mathbf{v} \cdot \mathbf{dS}.
   \label{eq:integral-advection-equation}
\end{equation}
Equation (\ref{eq:integral-advection-equation}) must hold for any
volume.  In particular, it must hold for every volume element
within the computational grid.  The exact integral relation is
then expressible in the following finite volume form for each grid
cell:
\begin{equation}
   \frac{\lambda^{\left( n+\mathrm{advection} \right)} -
         \lambda^{\left( n \right)}}{\Delta t} =
   - \frac{1}{\Delta V} \sum_{i = 1}^{6}
      \lambda_{i}^{\ast} \mathbf{v} \cdot \mathbf{\Delta S_{i}},
   \label{eq:finite-volume-advection-equation}
\end{equation}
where the summation is over all six faces on the surface of the
three-dimensional cell. The surface elements, $\mathbf{\Delta
S_{i}}$, are naturally face-centered with respect to the control
volume in question, so averages must be taken to obtain the
advection velocity components necessary to perform the dot product
for the momentum densities as shown in eq.\
(\ref{eq:finite-volume-advection-equation}).  We use second-order
accurate, linear averages to construct the advection velocities in
this case.  The amount of $\Lambda$ advected through each face is
given by an upwind biased, linear interpolation of the
distribution of $\lambda$ to give $\lambda^{\ast}$ as described by
\citet{vanleer79}.  By construction, the amount of $\Lambda$ that
is transported out of one cell immediately flows into the
neighboring cell; thus ensuring the conservative nature of the
advection scheme.

Unlike the ZEUS code, we do not use operator splitting along the
three separate dimensions during the advection step. Instead, we
perform the updates due to advection in all three dimensions
simultaneously. We thus avoid concerns about bias that may be
introduced by using an unsymmetrized ordering of the advection
sweeps.  A discussion of how we obtain second-order accuracy in
time for the advection step through time centering of the terms
appearing in eq.\  (\ref{eq:finite-volume-advection-equation}) is
presented in \S \ref{sec:time-centering}.

Our advection scheme automatically reverts to a first-order
accurate (upwind) scheme at local extrema in the primary fluid
variables.  In addition, it is necessary to introduce an
artificial viscosity to stabilize the scheme in the presence of
shocks. The artificial viscosity prescription we have implemented
is detailed in \S \ref{sec:artificial-viscosity}.

\subsection{Treatment of Source Terms}

The Lagrangian source terms for the momenta that are shown on the
right-hand sides of eqs.\
(\ref{eq:radial-euler-eq})-(\ref{eq:angular-euler-eq}) arise from
the forces of pressure and gravity, as well as from the
differentiation of the curvilinear basis vectors and the rotation
of the reference frame.  We have found it advantageous to combine
the pressure gradient with the gradient of the gravitational
potential, which results in a gradient of the sum of $H$ and
$\Phi$. Since the centrifugal force can also be expressed as the
gradient of a potential, it is included as well to form an
effective potential as defined in eq.\
(\ref{eq:effective-potential}).  As explained in \S \ref{sec:scf},
our initial models have the property that $\Phi^{\mathrm{eff}} =
\mathrm{constant}$ everywhere, hence to reasonably high precision
$\mathbf{\nabla} \Phi^{\mathrm{eff}} = 0$ throughout both stars
initially.

The expressions we have used for the source term updates of the
momentum densities are given here by expressions
(\ref{eq:s_update})-(\ref{eq:a_update}).  As with the advection
step, we do not use an operator splitting technique to evaluate
the source terms along the three separate coordinate dimensions;
instead, at each cell location, all updates due to Lagrangian
source terms are performed simultaneously.
\begin{eqnarray}
      \label{eq:s_update}
 & &     \frac{S^{\left( n + \mathrm{source} \right)}_{i,j+1/2,k+1/2} - 
            S^{\left( n \right)}_{i,j+1/2,k+1/2}}{\Delta t}  = \\
 & - &  \frac{\hat{\rho}_{i,j+1/2,k+1/2}}{\Delta R}
   \left[ \Phi^{\mathrm{eff}}_{i+1/2,j+1/2,k+1/2} - 
          \Phi^{\mathrm{eff}}_{i-1/2,j+1/2,k+1/2} \right] \nonumber  \\
 & + & \frac{\left( \hat{A}_{i,j+1/2,k+1/2} \right)^{2}}
               {\hat{\rho}_{i,j+1/2,k+1/2} R^{3}_{i}}
    +  \frac{2 \Omega \hat{A}_{i,j+1/2,k+1/2}}{R_{i}}; \nonumber
\end{eqnarray}
\begin{eqnarray}
   \label{eq:t_update}
 & &  \frac{T^{\left( n + \mathrm{source} \right)}_{i+1/2,j,k+1/2} -
         T^{\left( n \right)}_{i+1/2,j,k+1/2}}
        {\Delta t} = \\
 & - & \frac{\hat{\rho}_{i+1/2,j,k+1/2}}{\Delta z} \left[
         \Phi^{\mathrm{eff}}_{i+1/2,j+1/2,k+1/2} - \Phi^{\mathrm{eff}}_{i+1/2,j-1/2,k+1/2} \right]; \nonumber
\end{eqnarray}
\begin{eqnarray}
    \label{eq:a_update}
& &    \frac{A^{\left( n + \mathrm{source} \right)}_{i+1/2,j+1/2,k} -
          A^{\left( n \right)}_{i+1/2,j+1/2,k}}{\Delta t}  = \\
& - & \frac{\hat{\rho}_{i+1/2,j+1/2,k}}{\Delta \phi}
      \left[ \Phi^{\mathrm{eff}}_{i+1/2,j+1/2,k+1/2} -
             \Phi^{\mathrm{eff}}_{i+1/2,j+1/2,k-1/2} \right] \nonumber \\
& - & 2 \Omega \hat{S}_{i+1/2,j+1/2,k} R_{i+1/2}. \nonumber
\end{eqnarray}
Note that a caret identifies a variable whose value has been
interpolated to a spatial location that is different from the
variable's primary definition point as shown in Fig.\
\ref{fig:cell}.  These variables are given by volume-weighted
averages as follows:
\begin{eqnarray}
   & & \hat{A}_{i,j+1/2,k+1/2}  =  \frac{1}{4 R_{i}} \\
		       &   &   \left[ \left( A_{i+1/2,j+1/2,k} + A_{i+1/2,j+1/2,k+1} \right) \right. \nonumber \\
                       &   &   \left( R_{i} + \frac{1}{4} \Delta R \right) \nonumber \\
                       &   & + \left( A_{i-1/2,j+1/2,k} + A_{i-1/2,j+1/2,k+1} \right) \nonumber \\
                       &   &   \left.  \left( R_{i} - \frac{1}{4} \Delta R \right) \right], \nonumber \\
   & & \hat{S}_{i+1/2,j+1/2,k} =  \frac{1}{4 R_{i+1/2}} \\
		           &   &  \left[  \left( S_{i+1,j+1/2,k+1/2} + S_{i+1,j+1/2k-1/2} \right) \right. \nonumber \\
                           &   &          \left( R_{i+1/2} + \frac{1}{4} \Delta R_{i+1/2} \right) \nonumber \\
                           &   & +        \left( S_{i,j+1/2,k+1/2} + S_{i,j+1/2,k-1/2} \right) \nonumber \\
                           &   &   \left. \left( R_{i+1/2} - \frac{1}{4} \Delta R \right) \right], \nonumber \\
   & & \hat{\rho}_{i,j+1/2,k+1/2} = \frac{1}{2 R_{i}} \\
                              &   & \left[ \rho_{i+1/2,j+1/2,k+1/2}
                                    \left( R_{i} + \frac{1}{4} \Delta R \right) \right. \nonumber \\
                              &   & + \, \left. \rho_{i-1/2,j+1/2,k+1/2}
                                    \left( R_{i} - \frac{1}{4} \Delta R \right) \right], \nonumber \\
   & & \hat{\rho}_{i+1/2,j,k+1/2} = \frac{1}{2} \left( \rho_{i+1/2,j+1/2,k+1/2} \right. \\
                              &   & +            \left. \rho_{i+1/2,j-1/2,k+1/2} \right), \nonumber \\
   & & \hat{\rho}_{i+1/2,j+1/2,k} = \frac{1}{2}  \left( \rho_{i+1/2,j+1/2,k+1/2} \right.\\
                              &   & +             \left. \rho_{i+1/2,j+1/2,k-1/2} \right). \nonumber
\end{eqnarray}

\subsection{Artificial Viscosity}
\label{sec:artificial-viscosity} To stabilize the scheme in the
presence of shocks, we employ a planar, von Neumann artificial
viscosity that is active only for zones that are undergoing
compression. (See Stone \& Norman 1992 or Bowers \& Wilson 1991,
pg. 311 for more detailed discussions of artificial viscosity in
Eulerian hydrodynamics.) The momentum densities are updated from
the following finite-difference equations,
\begin{eqnarray}
& &   \frac{S^{\left( n + \mathrm{viscosity} \right)}_{i,j+1/2,k+1/2} -
            S^{\left( n \right)}_{i,j+1/2,k+1/2}}{\Delta t} = \frac{1}{\Delta R} \\
& & \left( Q^{RR}_{i+1/2,j+1/2,k+1/2} - Q^{RR}_{i-1/2,j+1/2,k+1/2} \right), \nonumber \\
& &   \frac{T^{\left( n + \mathrm{viscosity} \right)}_{i+1/2,j,k+1/2} -
            T^{\left( n \right)}_{i+1/2,j,k+1/2}}{\Delta t} = \frac{1}{\Delta z} \\
& &         \left( Q^{zz}_{i+1/2,j+1/2,k+1/2} - Q^{zz}_{i+1/2,j-1/2,k+1/2} \right), \nonumber \\
& &   \frac{A^{\left( n + \mathrm{viscosity} \right)}_{i+1/2,j+1/2,k} -
            A^{\left( n \right)}_{i+1/2,j+1/2.k}}{\Delta t} = \frac{1}{\Delta \phi} \\
& &         \left( Q^{\phi\phi}_{i+1/2,j+1/2,k+1/2} - Q^{\phi\phi}_{i+1/2,j+1/2,k-1/2} \right), \nonumber
\end{eqnarray}
where the diagonal components of the artificial viscosity are
given by,
\begin{eqnarray}
 Q^{RR}_{i+1/2,j+1/2,k+1/2} & = & \nu \rho_{i+1/2,j+1/2,k+1/2} \\
& &                                  \left( u_{i+1,j+1/2,k+1/2} - u_{i,j+1/2,k+1/2} \right)^{2}, \nonumber \\
 Q^{zz}_{i+1/2,j+1/2,k+1/2} & = & \nu \rho_{i+1/2,j+1/2,k+1/2} \\
& &                                   \left( w_{i+1/2,j+1,k+1/2} - w_{i+1/2,j,k+1/2} \right)^{2}, \nonumber \\
 Q^{\phi\phi}_{i+1/2,j+1/2,k+1/2} & = & \nu \rho_{i+1/2,j+1/2,k+1/2} \\
& &                                   \left( v_{i+1/2,j+1/2,k+1} - v_{i+1/2,j+1/2,k} \right)^{2}, \nonumber
\end{eqnarray}
if the velocity difference is negative; otherwise the components
of $Q$ are zero. Note that we neglect the shear components of
viscosity. The factor $\nu$ is a parameter that roughly dictates
the number of zones across which shock structures will be spread.
A value of $\nu = 2$ is typically sufficient. In keeping with our
overall adiabatic treatment of the flow (see \S
\ref{sec-analytic}), we neglect the generation of entropy by shock
compression.

In a binary system that is undergoing mass transfer, the accretion
stream will necessarily undergo a shock transition as it is
decelerated upon impact with the accreting star, or when it
intersects itself if the stream has sufficient angular momentum to
orbit the accretor.  In addition, even for a detached binary
simulation there will be weak standing shock fronts (as viewed in
the corotating frame of reference) at or near the surface of the
stars arising from the rapid deceleration of material falling onto
the stars.  We have found that even these weak shocks can have a
noticeable impact on the quality of the solution in long time
evolutions of detached systems unless artificial viscosity is used
to damp the resulting oscillations.

\subsection{Time Centering}
\label{sec:time-centering} The timestep cycle is split between the
application of source, advection and viscosity operators. First,
the source terms are applied for one half of a timestep.  Next,
all updates due to advection are performed for a full timestep and
the viscosity updates are applied to the momentum densities.
Finally, the second half of the source operators are applied.  The
source and advection steps are thereby staggered in time when
viewed over several iteration cycles for a constant value of the
timestep.

The advection is time-centered by first performing half a timestep
of fictitious advection in order to obtain ``time-centered''
velocities for constructing the face-centered advection velocity
components that appear in eq.\
(\ref{eq:finite-volume-advection-equation}). The full timestep of
advection is then performed.  The components of the viscosity
tensor are constructed from the velocity and density estimates at
the midpoint of the timestep as well.

Since the momentum densities themselves also appear in the source
terms of eqs.\ (\ref{eq:radial-euler-eq}) and
(\ref{eq:angular-euler-eq}), similar care must be taken with their
centering in time.  The source operators are applied in a
fictitious source step to obtain the angular and radial momentum
densities at a point half a timestep in the future.  These values
are then used to update the momentum densities through a full
timestep.  As the timestep value changes from iteration to
iteration, this algorithm for time centering the source terms is
not formally accurate to second order. However, in real
computations the character of the flow and, hence, the maximal
signal velocity do not change rapidly over the course of a
timestep cycle so that one may expect the resulting inaccuracies
in the time centering of the source terms to be small. The other
terms that appear in the source operators, including the
gravitational potential, are all calculated at the approximate
midpoint in time between the source steps.

\subsection{Timestep Formulation and Boundary Conditions}
Since we explicitly integrate the fluid equations in time, the
timestep is limited in size by the familiar
Courant-Friedrichs-Lewy (CFL) stability criterion which ensures
the time increment is small enough so that no characteristic can
cross a cell in a single timestep.  Specifically,
\begin{equation}
   \Delta t = \mathrm{min} \left[ \frac{\Delta R}{c + \left|u \right|},
                     \frac{\Delta Z}{c + \left|w \right|},
             \frac{R \Delta \phi}{c + \left|v \right|} \right],
\end{equation}
where $c$ is the speed of sound. In practice we limit the timestep
to a half the CFL time. Since we have introduced the diffusion
terms associated with artificial viscosity, the timestep must also
satisfy the condition, (see p.270 of Bowers \& Wilson 1991),
\begin{equation}
   \Delta t \leq \frac{1}{4} \mathrm{min} \left[  \frac{\rho \Delta R}{Q^{RR}},\frac{\rho \Delta Z}{Q^{ZZ}},
                                  \frac{\rho R \Delta \phi}{Q^{\phi\phi}} \right] ^{1/2}.
\end{equation}

The boundary conditions for the fluid variables at the external
boundaries are to allow the fluid to flow freely off the grid but
to not allow material to flow back from the outermost layer of
boundary cells.  The central annulus of cells that has an inner
radius at the coordinate axis is treated as a single azimuthally
averaged cell for each layer in the vertical direction.

\subsection{Parallelization of Hydrodynamics Algorithm}
As it is our intention to perform high resolution simulations, it
is imperative that the work load within the simulation be
distributed amongst many processors so that the simulations may be
conducted in a reasonable amount of time and not exceed the
available memory of a single node.  The fluid dynamics equations,
being hyperbolic partial differential equations, are ideally
suited to a simple domain decomposition or single program multiple
data (SPMD) parallelization model.  Each computational task
performs the same operations on their own block of the global data
arrays with communication only being necessary between nearest
neighbor tasks that share a boundary of ghost zones that is
one-cell thick (this ghost zone thickness is dictated by the order
of our advection and finite-difference operators).  We have
written the program in Fortran90 with explicit message passing
being performed with MPI (Message Passing Interface) subroutine
calls.  The resulting parallel code performance scales linearly
with the number of processors for 4 to 128 processors on the Cray
T3E.  Similar behavior is also seen on the IBM SP platform.

\subsection{Solution of Poisson's Equation}

We are seeking to solve Poisson's equation for an isolated
distribution of mass. The correct boundary condition in this
instance is that the potential goes to zero at infinity.  As we
only construct the solution on a finite domain we must specify the
gravitational potential (or its gradient) on some boundary that
encloses all the mass in the simulation.  We construct the
boundary potential using a novel technique based on a compact
representation of the cylindrical Greens function in terms of
half-integer degree Legendre functions of the second kind as
described by \citet{cohl99}.  The boundary potential is then
simply given by the convolution of the appropriate Greens function
with the density distribution.  This method is capable of
generating the exact solution for a discretized mass distribution
and has the attractive feature that it can be applied to very
flattened bodies without suffering penalties in either performance
or accuracy.

In order to obtain the interior solution for the gravitational
potential, Poisson's equation is first Fourier transformed in the
azimuthal direction, then the resulting set of two-dimensional
partial differential equations (Helmholtz equations) for the
decoupled Fourier amplitudes are solved using an alternating
direction implicit (ADI) scheme
\citep[c.f.,][]{peaceman55,black75}. The solution is then
transformed back to real space.

The solution of Poisson's equation requires special care in the
context of parallel computing because the solution necessarily
involves global communication as the character of the underlying
physical law is action at a distance.  The algorithms we have
implemented for computing the gravitational potential are well
suited to a cylindrical geometry and very efficient in a
distributed computing environment.  Parallel communications are
used to transpose the data so that all the data in a given
dimension are in local memory at one time. When operations are to
be performed along a different dimension, the data are transposed
again. This allows us to send a relatively few number of large
messages. Further details regarding our solution of Poisson's
equation in a parallel computing environment can be found in
\citet{cohl97}.

\section{Test Cases}

Here we present results from three different types of tests that
we have used to evaluate and quantify the accuracy of our
computational tools.  In all tests we compare a known, although
not necessarily analytical, solution with the calculated numerical
solution.

\subsection{Riemann Shock Tube Test}
\label{sec:shock-test}

As a check of the stability of our code in the presence of,
ideally, discontinuous jumps in the fluid variables we have solved
Sod's shock tube problem \citep{sod78} with the initial
discontinuity lying along a plane of constant $z = z_{0}$. Sod's
shock tube problem  presents a useful hydrodynamic test because
the solution is known analytically and contains the three simple
waves that can occur in ideal fluid flow.  Of these simple waves,
it is the shock wave that concerns us most.  Our goal is not to
resolve the details of the shock structure but rather to ensure
that our algorithm is well behaved (numerically stable and yields
an acceptable solution) in the presence of shocks.

The initial conditions for Sod's shock tube problem are that the
velocity, $\mathbf{v}$, is zero everywhere; for $z \leq z_{0}$,
the pressure, density and internal energy per unit mass take on
the values $P_{l} = 1.0$, $\rho_{l} = 1.0$, $\epsilon_{l} = 2.5$;
for $z > z_{0}$, $P_{u} = 0.1$, $\rho_{u} = 0.125$, $\epsilon_{u}
= 2.0$.  The fluid flow is characterized by an adiabatic exponent,
$\gamma = 1.4$.

\begin{figure}
   \plotone{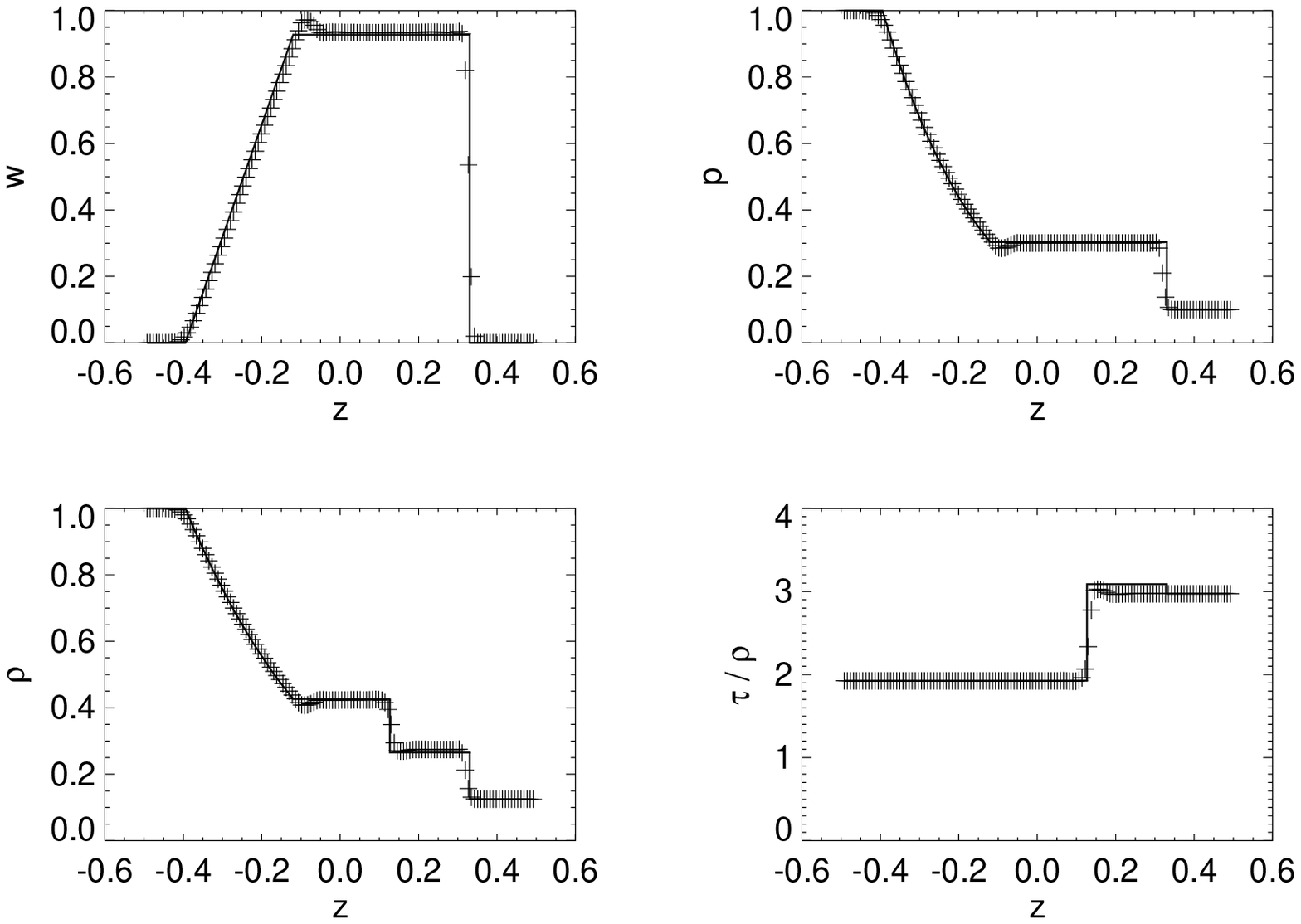}
   \figcaption[f5.eps]{Clockwise from the upper left panel, the vertical velocity, pressure,
   the ratio of the entropy tracer to the mass density
   and the mass density are shown as a function of position $z$ at time $t = 0.247$ for Sod's shock tube problem.
   The initial discontinuity was placed at $z = -0.1$.  The computed solution for a randomly chosen column
   of cells at constant radius and azimuth is plotted as crosses;
   the solid curves are the analytic solution.\label{fig:sod_shock_tube}}
\end{figure}

The computed solution for the vertical velocity, $w$, pressure,
$p$, mass density $\rho$, and the quantity, $ \tau / \rho$ (which
is proportional to the polytropic constant and, hence, the
entropy; see eq.\ \ref{eq:entropy}) is plotted along with the
analytical solution at time $t = 0.247$ in Fig.\
\ref{fig:sod_shock_tube}. The computed points are not average
values but are instead the values for a random column of cells at
constant radius and azimuth within the three-dimensional grid. The
calculation was performed with a coefficient for the artificial
viscosity of $ \nu = 2.0 $ and with 130 vertical zones. The
initial discontinuity was placed at $z_{0} = -0.1$ and the grid
extended from $- 1 / 2$ to $1 / 2$ in the vertical
direction.

The results from this simulation compare favorably to the results
produced by other second-order accurate, Eulerian hydrodynamics
programs with artificial viscosity
\citep[c.f.,][]{hawley84,stone92,lufkin93}. The shock front is
spread out over approximately three zones, and there is no
indication of numerical instability in the solution for the
shocked gas.  The contact discontinuity is likewise spread out
over about three zones due to the numerical diffusion inherent in
a second-order accurate Eulerian scheme.  There is some
disagreement between the computed and analytical solution at the
tail of the rarefaction wave.  This phenomenon has been
investigated by \citet{norman83} and results from an inconsistent
representation of the analytic viscous equations in finite
difference form.  Finally, immediately behind the shock, the
analytical solution for ${\tau}/{\rho}$ disagrees slightly with
the computed solution for the shocked gas.  This is because the
analytical solution includes a small increase in the entropy of
the fluid that passes through the shock whereas, as discussed in
\S \ref{sec:artificial-viscosity}, we have elected to treat all of
the flow as through it is adiabatic, that is, by setting $ds/dt =
0$ in eq. (33).  [Agreement with the analytical result could
readily have been achieved by constructing an appropriate
expression for $ds/dt$ in terms of the artificial viscosity in
order to account for dissipation in the shock, as shown for
example in eqs. (11) and (37) of Stone \& Norman 1992.] In effect,
we have assumed that each fluid element that passes through the
shock is immediately able to cool back down to a temperature that
places it back on its original pre-shock adiabat.

We note that we have used the gradient of the pressure as opposed
to the density times the gradient of the enthalpy for the solution
of Sod's problem.  Due to the pathological nature of the
discontinuous initial conditions, a correct solution cannot be
obtained if the enthalpy term is used with our chosen centering of
the fluid variables.

\subsection{Test of Poisson Solver}

\citet{cohl99} have published exhaustive tests showing the
accuracy with which we are able to evaluate the gravitational
potential on the boundary of our cylindrical coordinate grid.  In
order to ascertain the accuracy with which we are able to
determine the force of gravity arising from the fluid everywhere
inside the grid, we have calculated the potential and its
derivatives for a uniform-density sphere of radius $R_{\ast}$ and
density $\rho_{\ast}$, centered at an arbitrary position on the
grid, $\mathbf{r}_{0}$. The analytical potential is
\begin{equation}
   \Phi \left( \mathbf{r} \right) = \left\{
              \begin{array}{ll}
                 - 2 \pi G \rho_{\ast} \left( R^{2}_{\ast} - \displaystyle{\frac{d^{2}}{3}} \right) &
              d < R_{\ast} \\
         - \displaystyle{\frac{4 \pi}{3}} G \rho_{\ast} \displaystyle{\frac{R^{3}_{\ast}}{d}} &
               d > R_{\ast},
          \end{array}
          \right.
\end{equation}
where $d = \left| \mathbf{r} - \mathbf{r}_{0} \right|$.  The
corresponding derivatives appearing in the gravitational force
are:
\begin{eqnarray}
 \frac{\partial \Phi}{\partial R} & = &
          \frac{4 \pi G \rho_{\ast}}{3} \left[ \left( R \cos{\phi} - x_{0} \right) \cos{\phi} \right. \nonumber \\
& + & \left. \left( R \sin{\phi} - y_{0} \right) \sin{\phi} \right], \nonumber   \\
  \frac{\partial \Phi}{\partial z} & = &
                \frac{4 \pi G \rho_{\ast}}{3} \left( z - z_{0} \right), \\
  \frac{\partial \Phi}{\partial \phi} & = &
                \frac{4 \pi G \rho_{\ast}}{3} \left[ -R  \left( R \cos{\phi} - x_{0} \right) \sin{\phi} \right. \nonumber \\
& + & \left. R \left( R \sin{\phi} - y_{0} \right) \cos{\phi} \right], \nonumber
\end{eqnarray}
for $d < R_{\ast}$, and
\begin{eqnarray}
  \frac{\partial \Phi}{\partial R} & = &
          \frac{4 \pi G \rho_{\ast}}{3} \frac{R^{3}_{\ast}}{d^{3}}
      \left[ \left( R \cos{\phi} - x_{0} \right) \cos{\phi} \right. \nonumber \\
& + & \left. \left( R \sin{\phi} - y_{0} \right) \sin{\phi} \right], \nonumber \\
  \frac{\partial \Phi}{\partial z} & = &
          \frac{4 \pi G \rho_{\ast}}{3} \frac{R^{3}_{\ast}} {d^{3}}
      \left( z - z_{0} \right), \\
  \frac{\partial \Phi}{\partial \phi} & = &
          \frac{4 \pi G \rho_{\ast}}{3} \frac{R^{3}_{\ast}}{d^{3} }
          \left[ - R \left( R \cos{\phi} - x_{0} \right) \sin{\phi} \nonumber \right. \\
& + &  \left. R \left( R \sin{\phi} - y_{0} \right) \cos{\phi} \right], \nonumber
\end{eqnarray}
for  $d > R_{\ast}$.

In Table \ref{tab:poisson_rel_err} we present the average relative
error in the potential and the average absolute error in the three
derivatives for a uniform density sphere ($\rho_{\ast} = 1$) of
radius $R_{\ast} = 1 / 3$ placed at the origin and at $x_{0}
= 1 / 2$ for a representative set of grid resolutions.  The
grid extends from $0$ to $1$ in radius and from $- 1 / 2$ to
$1 / 2$ in the vertical direction. Similarly, in Table
\ref{tab:poisson_max_err} we present the maximum errors for the
same quantities.

\begin{deluxetable}{ccccllll}
   \tablenum{4}
   \tablewidth{0pt}
   \tablecolumns{8}
   \tablecaption{Average Error for Gravitational Potential and Force\label{tab:poisson_rel_err}}
   \tablehead{
     \colhead{Origin} & \colhead{R} & \colhead{z} & \colhead{$\phi$} & \colhead{$\Phi$} &
      \colhead{$\partial_{R} \Phi$} & \colhead{$\partial_{z} \Phi$} & \colhead{$\partial_{\phi} \Phi$} }
   \startdata
                     0 & 66  & 66  & 64  & $1.0 \times 10^{-2}$ & $3.4 \times 10^{-3}$ &
                                           $2.7 \times 10^{-3}$ & $4.0 \times 10^{-18}$  \\
                       & 66  & 66  & 128 & $1.0 \times 10^{-2}$ & $3.4 \times 10^{-3}$ &
                                           $2.7 \times 10^{-3}$ & $4.0 \times 10^{-18}$   \\
	               & 130 & 130 & 128 & $3.1 \times 10^{-3}$ & $1.1 \times 10^{-3}$ &
	                                   $9.7 \times 10^{-4}$ & $4.2 \times 10^{-18}$   \\
	               & 130 & 130 & 256 & $3.1 \times 10^{-3}$ & $1.1 \times 10^{-3}$ &
	                                   $9.7 \times 10^{-4}$ & $4.2 \times 10^{-18}$   \\
$0.5 \hat{\mathbf{x}}$ & 66  & 66  & 64  & $3.3 \times 10^{-3}$ & $9.5 \times 10^{-4}$ &
                                           $7.3 \times 10^{-4}$ & $3.3 \times 10^{-4}$    \\
                       & 66  & 66  & 128 & $2.6 \times 10^{-4}$ & $3.6 \times 10^{-4}$ &
                                           $3.6 \times 10^{-4}$ & $1.4 \times 10^{-4}$     \\
	               & 130 & 130 & 128 & $2.3 \times 10^{-4}$ & $2.3 \times 10^{-4}$ &
	                                   $1.8 \times 10^{-4}$ & $7.0 \times 10^{-5}$     \\
	               & 130 & 130 & 256 & $9.9 \times 10^{-5}$ & $8.6 \times 10^{-5}$ &
	                                   $7.1 \times 10^{-5}$ & $3.6 \times 10^{-5}$
   \enddata
\end{deluxetable}

\begin{deluxetable}{ccccllll}
   \tablenum{5}
   \tablewidth{0pt}
   \tablecolumns{8}
   \tablecaption{Maximum Error for Gravitational Potential and Force\label{tab:poisson_max_err}}
   \tablehead{
            \colhead{Origin} & \colhead{R} & \colhead{z} & \colhead{$\phi$} & \colhead{$\Phi$} &
             \colhead{$\partial_{R} \Phi$} & \colhead{$\partial_{z} \Phi$}  & \colhead{$\partial_{\phi} \Phi$} }
   \startdata
                             0 & 66  & 66  & 64  & $1.3 \times 10^{-2}$ & $4.7 \times 10^{-2}$ &
                                                   $4.8 \times 10^{-2}$ & $8.3 \times 10^{-17}$   \\
		               & 66  & 66  & 128 & $1.3 \times 10^{-2}$ & $4.7 \times 10^{-2}$ &
		                                   $4.8 \times 10^{-2}$ & $7.7 \times 10^{- 17}$   \\
		               & 130 & 130 & 128 & $4.6 \times 10^{-3}$ & $2.3 \times 10^{-2}$ &
		                                   $2.1 \times 10^{-2}$ & $6.4 \times 10^{-17}$    \\
		               & 130 & 130 & 256 & $4.6 \times 10^{-3}$ & $2.3 \times 10^{-2}$ &
		                                   $2.1 \times 10^{-2}$ & $7.3 \times 10^{-17}$    \\
        $0.5 \hat{\mathbf{x}}$ & 66  & 66  & 64  & $8.8 \times 10^{-3}$ & $1.2 \times 10^{-1}$ &
                                                   $1.1 \times 10^{-1}$ & $2.4 \times 10^{-2}$     \\
                               & 66  & 66  & 128 & $3.7 \times 10^{-3}$ & $5.0 \times 10^{-2}$ &
		                                   $5.2 \times 10^{-2}$ & $2.2 \times 10^{-2}$     \\
                               & 130 & 130 & 128 & $3.4 \times 10^{-3}$ & $6.0 \times 10^{-2}$ &
                                                   $4.7 \times 10^{-2}$ & $1.3 \times 10^{-2}$     \\
	                       & 130 & 130 & 256 & $1.0 \times 10^{-3}$ & $3.3 \times 10^{-2}$ &
	                                           $3.2 \times 10^{-2}$ & $1.8 \times 10^{-2}$
   \enddata
\end{deluxetable}

The region near the surface of the sphere contains the largest
errors in the potential solution \citep[c.f.,][]{stone92}. At the
surface, the density falls discontinuously to zero and the slope
of the solution changes abruptly.  When placed at the origin, this
high error region is resolved by a larger number of smaller-volume
cells than when the sphere is placed off-axis in the grid.  This
results in a worse average error for the potential and its radial
and vertical derivatives for the axisymmetric solution despite the
fact that the maximal errors are generally smaller in this
instance.

For the case where the sphere is centered on the origin of the
computational grid, the resulting potential is axisymmetric to
machine accuracy.  The average relative error in the potential and
the average absolute error in the radial and vertical derivatives
all decrease by a factor of about three as the radial and vertical
resolutions are doubled. As expected for an axisymmetric mass
distribution the quality of the solution is independent of the
number of azimuthal zones.  The maximum values of the relative
error in the potential decrease by a factor of about three as well
and the maximum value in the absolute error of the radial and
vertical derivatives has been cut in half as the number of radial
and vertical zones doubles.

When the sphere is placed off axis, the convergence pattern is
much more difficult to recognize. For the off-axis test at the
highest resolution (the same radial and azimuthal resolution that
we currently use for binary evolutions), we are able to obtain a
solution that is accurate to one part in $10^{4}$, on average, for
the potential.  Similarly, the finite-difference and analytical
components of the derivatives of the potential agree to better
than 4 decimal places on average.

\subsection{Test of Hydrostatic Equilibrium}
\label{sec:equilibrium_test}

A stringent test of our coupled solution of Poisson's equation and the fluid dynamical
equations --- and one that may seem trivial at first mention --- is how well we are able to
maintain hydrostatic equilibrium for a simple system such as a spherical polytrope
that is placed off axis in the grid.  While our hydrodynamics implementation is
conservative with respect to the advection of the fluid, there is no guarantee
that the total momentum is conserved once the action of the Lagrangian source
terms are included.  Throughout a mass-transfer simulation, the bulk of the fluid
should remain near hydrostatic equilibrium and the correct response of both components
to their changing mass can be limited by the accuracy to which force balance is
maintained.

To perform this test, we have placed a spherical, $n =
3 / 2$ polytrope of radius $R = 0.38$ in a cylindrical grid
of total radius 1.0, but with a variety of different resolutions.
The polytrope is centered at $x \approx 0.58$.  In each case, the
initial density distribution was generated with our SCF code (with
only one star present and no frame rotation), and the initial
velocities were zero everywhere.  Using our full gravitational
hydrodynamics code, we then permitted the fluid system to evolve
in time.

\begin{figure}
   \plotone{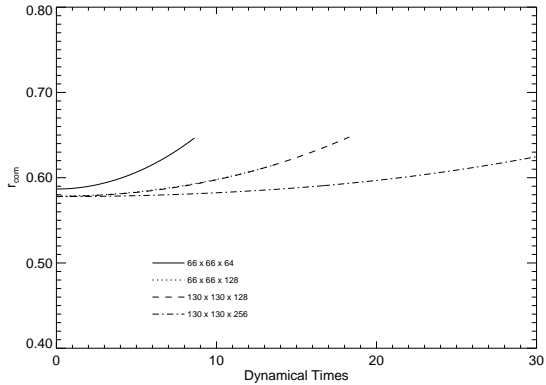}
   \figcaption[f6.eps]{Distance from origin to the center of mass as a function of time
   (measured in dynamical times) for the off-axis spherical polytrope described in \S \ref{sec:equilibrium_test}.
   The different curves correspond to calculations performed at the indicated resolution, in terms
   of the number of radial by vertical by azimuthal zones.
   The curves representing the simulations at resolutions of $66 \times 66 \times 128$ and
   $130 \times 130 \times 128$ lie on top of one another. \label{fig:lonestar_com}}
\end{figure}

Over the course of the evolutions, each isolated star drifts
outwards as if acted on by a constant force.  This drift is shown
in Fig.\ \ref{fig:lonestar_com} where we have plotted the location
of the center of mass of the star as a function of time for grids
of varying resolution. We have normalized the evolution time to
the dynamical time as given by, for example,
\citet{chandrasekhar39}.  Specifically, $t_{\mathrm{dynamical}} =
\sqrt{ \left( 3 \pi \right) / \left( 16 G \bar{\rho} \right)}$, and for an $n =
3 / 2$ polytrope with central density of unity the average
density is $\bar{\rho} = 0.1669$. The size and rate of the drift
decreases as the azimuthal resolution increases.

\begin{figure}
   \plotone{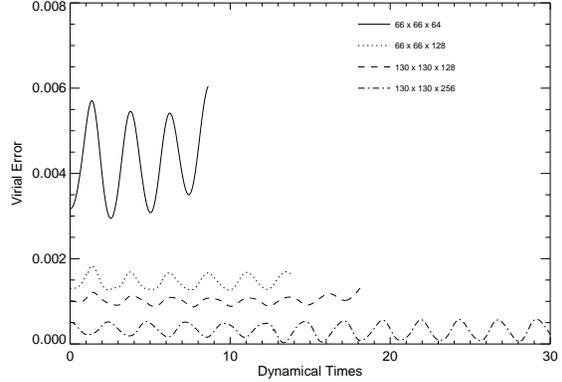}
   \figcaption[f7.eps]{The virial error, as given by eq.\  (\ref{eq:mod_virial-error}) 
   ( the virial error with the kinetic energy term omitted) is plotted as a function of 
   the number of dynamical times for the off-axis spherical polytrope described in 
   \S \ref{sec:equilibrium_test}.  The meaning of the different curves is the same as 
   in Fig.\  \ref{fig:lonestar_com}.\label{fig:lonestar_virial}}
\end{figure}

\begin{figure}
   \plotone{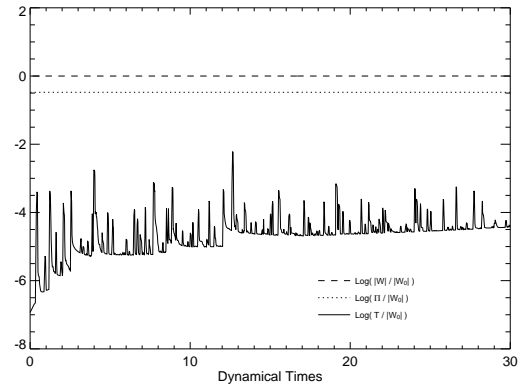}
   \figcaption[f8.eps]{The logarithm of the three components to the virial error as 
   given in eqs.\ (\ref{eq:t})--(\ref{eq:pi}) normalized to the initial value of the 
   gravitational potential energy is plotted as a function of time for the off-axis 
   spherical polytrope evolved in a grid of resolution $130 \times 130 \times 256$.  
   \label{fig:lonestar_virial_components}} 
\end{figure}

As another measure of the quality of the steady state equilibrium
from these spherical polytropes we show a modified virial error,
\begin{equation}
   VE = \frac{ \left( W + 3 \Pi \right) }{ | W | },
   \label{eq:mod_virial-error}
\end{equation}
in Fig.\ \ref{fig:lonestar_virial}.  This differs from the
definition given in eq.\ (\ref{eq:virial-error}) in that we have
neglected the kinetic energy term, $K$.  As can be seen in Fig.\
\ref{fig:lonestar_virial_components} where we show the log of $| W
|$, $\Pi$ and $K$ normalized to the initial value of $| W |$ for
the highest resolution simulation (computed with 130 radial and
vertical zones by 256 azimuthal zones), some peaks in kinetic
energy, which are noise, are of approximately the same size as the
sum of $W$ and $3 \Pi$ in spite of the fact that the kinetic
energy is insignificant compared to either the thermal or
gravitational energies. Overall, the virial error decreases by a
factor of approximately 6 from the lowest to highest resolution
simulation.  At the highest resolution presented the virial error
is $0.05 \%$ and the polytrope oscillates with amplitude of
approximately $0.02 \%$ for 30 dynamical times.  This shows that
the isolated star remains in hydrostatic equilibrium to a very
high degree of accuracy, even when placed off-axis in our
computational grid.

There is no significant improvement in the drift of the system
center of mass when only the radial and vertical resolutions are
increased. (In Fig.\ \ref{fig:lonestar_com}, compare the curves
for the simulations at resolutions of 66 radial, 66 vertical by
128 azimuthal zones and 130 radial, 130 vertical by 128 azimuthal
zones.) But there is an improvement in the virial error between
these two simulations.  This suggests that there are two limiting
numerical effects in play.  One dictates the resolution of the
equilibrium state itself; the other causes a displacement of that
equilibrium state.  The former effect converges with the
finite-difference size isotropically, while the latter depends
only on the azimuthal resolution. When trying to resolve a highly
nonaxisymmetric object, such as an off-axis sphere, within a
uniform cylindrical coordinate grid, different parts of the star
are resolved to varying degrees and it is not surprising that the
convergence of the numerical solution is not describable in simple
terms.

\section{Benchmark Simulations}

In this section we present results from two simulations of
detached binaries that we have performed to ascertain the
precision with which we can expect to carry out future simulations
of semi-detached binary systems (systems undergoing mass
transfer). One binary is an equal mass system with identical
components (see Model 5 in Table \ref{tab:scf_models} and Figs.\
\ref{fig:scf_side}--\ref{fig:scf_top}; hereafter referred to as
the EB system) and the other system has a mass ratio $q = 0.8436$
(see Model 6 in Table \ref{tab:scf_models} and Figs.\
\ref{fig:scf_side}-- \ref{fig:scf_top}; hereafter referred to as
unequal binary or UB system). The EB system was constructed to
resemble the single star used for the test of hydrostatic
equilibrium in \S \ref{sec:equilibrium_test}.  This enables us to
compare the systematic errors in the case of a binary system given
the errors observed when only gravity, pressure and the curvature
force came into play. Each component of the EB system differs from
the isolated, spherical star in that each is flattened by the
synchronous rotation of the system and tidally distorted by its
companion, but the components have a comparable size, in terms of
grid cells, and the same central density and polytropic index as
the isolated sphere.

Previous simulations of equal-mass barotropic stars have shown
that it is important to conduct the evolutions in a frame of
reference that renders the binary as close to static as possible
in order to minimize the effects of numerical diffusion arising
from the finite accuracy of Eulerian advection schemes (New \&
Tohline 1997; SWC). With this in mind, our EB and UB simulations
have been conducted in a frame of reference rotating with the
orbital angular velocity $\Omega$ of the system, as obtained by
our SCF technique.

In dealing with unequal-mass systems we have discovered another
subtle, but important issue that should be addressed with care
when ``transporting'' an initial hydrostatic model from the grid
of the SCF code into the grid of the hydrodynamics code.  During
each SCF iteration, the system's center of mass is not fixed to
any location beyond the fact that, by symmetry, it must lie along
the line of centers.  In general, then, we must translate the
density field as we introduce it into the hydrocode so that the
system center of mass coincides with the $z$-axis, which is taken
to be the rotation axis for the hydrodynamic evolution.  If we
could perform this translation perfectly, all initial fluid
velocities would be identically zero relative to the hydrodynamic
reference frame.  Because of the inherent symmetry of an
equal-mass binary system, this was in fact the case for our EB
system by construction.  For the UB system, however, the center of
mass of our converged SCF model was displaced by a small distance
from the rotation axis.  Specifically, $\mathbf{R_{com}} = 2.842
\times 10^{-6} \mathbf{\hat{x}}$, which corresponded to only $4
\times 10^{-4} \Delta R$, where $\Delta R$ is the radial extent of
each grid cell. As we introduced the SCF model into the
hydrodynamical grid, we therefore also ascribed nonzero velocities
as initial conditions according to the relation,
\begin{equation}
   \mathbf{v} = - \mathbf{\Omega} \times \mathbf{R_{com}}.
   \label{eq:com_vel}
\end{equation}
Because the displacement $\mathbf{R_{com}}$ was quite small for
our UB system, the initial velocities prescribed through eq.\
(\ref{eq:com_vel}) were also very small. Nevertheless, it was
necessary to include them in order to achieve the best possible
steady-state configurations corresponding to the stars following
circular orbits. This implies a uniform initial velocity for the
system (see further discussion below).

After the binary models were introduced into the hydrodynamics code,
both were evolved through more than five orbits.  (See the first row
of Table \ref{tab:systems}, where the total evolution time for both
simulations is tabulated in units of each system's orbital period P.)
As is recorded in the last three rows of Table \ref{tab:systems},
the EB system was run on 64 nodes
of a Cray T3E 600 for a total of 173 wall-clock hours (that is,
the simulation required on average 2409 processor-hours per orbit)
and the UB system was run on 8 dual processor nodes of an IBM SP 3
for a total of 265 wall-clock hours (that is, the simulation required
on average 819 processor-hours per orbit).
Many different diagnostic parameters were followed throughout both
evolutions in order to assess the quality of the initial SCF models
and to determine with what accuracy the hydrodynamical equations were
being integrated forward in time.   In the following paragraphs, we
present the time-evolutionary behavior of a number of these key physical
parameters.

\begin{deluxetable}{ccc}
   \tablenum{6}
   \tablewidth{0pt}
   \tablecolumns{3}
   \tablecaption{Quantities of Interest for Benchmark Simulations\label{tab:systems}}
   \tablehead{
             \colhead{Quantity} & \colhead{Equal Mass Binary (EB)} & \colhead{Unequal Mass Binary (UB)} }

   \startdata
         $\frac{t}{P}$                                    & 5.314                      & 5.178 \\
         $\frac{\Delta M_{\mathrm{1}}}{M}$                & $-9.0 \times 10^{-6}$      & $-3.0 \times 10^{-5}$ \\
         $\frac{\Delta M_{\mathrm{2}}}{M}$                & $-1.0 \times 10^{-5}$      & $-1.1 \times 10^{-6}$ \\
         $\frac{\Delta M}{M}$                             & $-1.9 \times 10^{-5}$      & $-1.4 \times 10^{-5}$ \\
 $\left( \frac{\Delta a}{a} \right)_{\mathrm{secular}}$   & $-2.9 \times 10^{-4}$      & $-1.9 \times 10^{-4}$ \\
 $\left( \frac{\Delta a}{a} \right)_{\mathrm{epicyclic}}$ &  $5.0 \times 10^{-4}$      &  $2.2 \times 10^{-4}$\\
          $\frac{\Delta J_{z}}{J_{z}}$                    & $+1.1 \times 10^{-4}$      & $+1.5 \times 10^{-4}$  \\
          Machine                                         & Cray T3E 600               & IBM SP3 \\
          Processors                                      & 64                         & 16 \\
	  $T_{\mathrm{Wall Clock}}$                       & 173 hours                  & 265 hours
    \enddata
\end{deluxetable}

\subsection{Stars in Hydrostatic Balance}

\begin{figure}
    \plotone{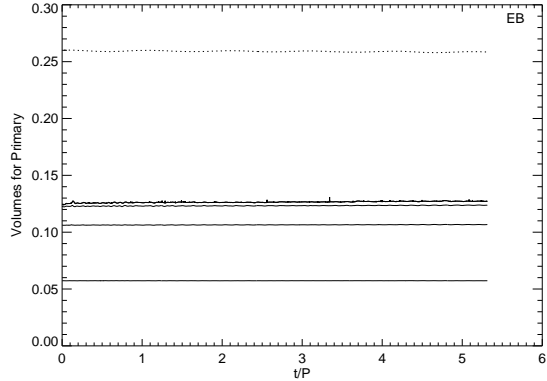}
    \figcaption[f9.eps]{The Roche volume (dashed curve) and volume occupied by material
    more dense than $10^{-1}$, $10^{-2}$, $10^{-3}$, $10^{-4}$ and $10^{-5}$
    (solid curves from bottom to top) are plotted as a function of the orbital time 
    for one component of the EB system.\label{fig:eb_vols}}
\end{figure}

\begin{figure}
    \plotone{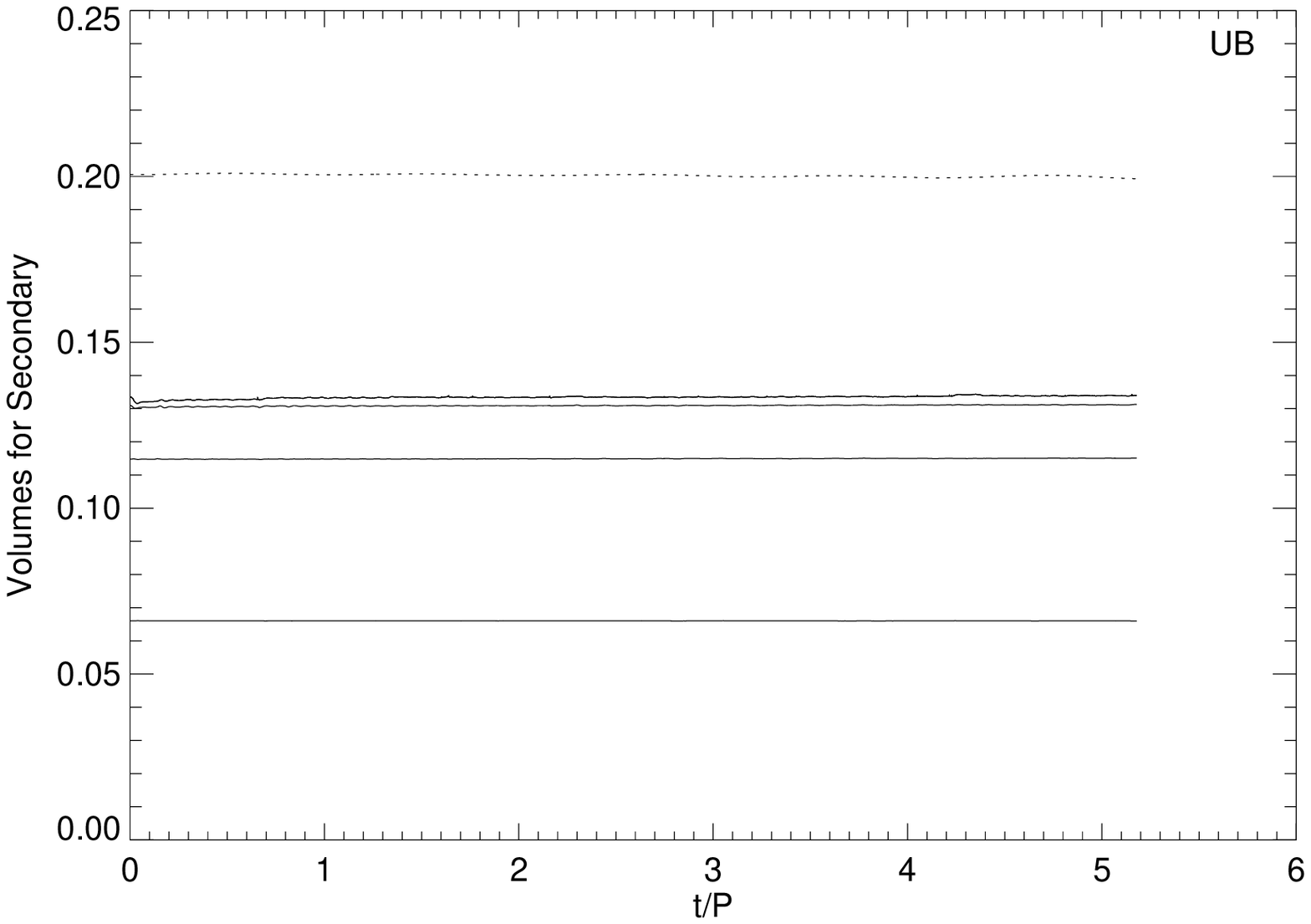}
    \figcaption[f10.eps]{The Roche volume (dashed curve) and volume occupied by material 
    more dense than $10^{-1}$, $10^{-2}$, $10^{-3}$, $10^{-4}$ and $10^{-5}$ 
    (solid curves from bottom to top) as a function of the orbital time for the 
    secondary component of the UB system.  \label{fig:ub_1_vols}}
\end{figure}

\begin{figure}
    \plotone{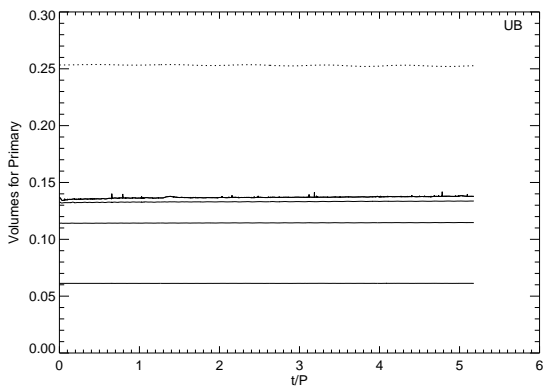}
    \figcaption[f11.eps]{The Roche volume (dashed curve) and volume occupied by material 
    more dense than $10^{-5}$, $10^{-4}$, $10^{-3}$, $10^{-2}$ and $10^{-1}$ 
    (solid curves from bottom to top) as a function of the orbital time for the 
    primary component of the UB system.  \label{fig:ub_2_vols}}
\end{figure}

Throughout both evolutions, the individual stellar components were
largely static and remained well within their respective Roche
lobes. In an effort to illustrate this, Fig.\ \ref{fig:eb_vols}
shows as a function of time the computed Roche lobe volume (dashed
curve) and the volumes (solid curves) occupied by material more
dense than $10^{-1}$, $10^{-2}$, $10^{-3}$, $10^{-4}$, and $10^{-5}$ for one
component of the EB system. (For reference, the initial SCF
density fields have values of a few times $10^{-5}$ at the edge of
the stars; see the isodensity contours drawn in Figs.\
\ref{fig:scf_side} and \ref{fig:scf_top}.) The same information is
plotted in Figs.\  \ref{fig:ub_1_vols} and \ref{fig:ub_2_vols} for
the secondary and primary components, respectively, of the UB
system.  These figures illustrate that the rotationally flattened
and tidally distorted models generated by our SCF code exhibit
excellent detailed force balance throughout their
three-dimensional structures, and that there is an excellent match
between the algorithmic expressions that determine an equilibrium
state in the SCF code and force balance in the hydrodynamics code.

\subsection{Mass Conservation}

In an effort to determine how well mass is conserved throughout
an evolution for each star, individually, as well as for the system
as a whole, we tracked three separate volume integrals over the
mass density: $M_1$, defined as the mass bound to the
primary; $M_2$, defined as the mass bound to the secondary;
and $M_{\rm envelope}$, defined as the mass that lies outside of both
stars but inside the boundaries of the computational
grid.  As is illustrated by frames 5 and 6 of Figs.\
\ref{fig:scf_side} and \ref{fig:scf_top}, in the initial state
it is easy to evaluate these three integrals because the
edges of the two stars are well-defined.
Specifically, when normalized to each
system's total mass, $M_1 = (1 + q)^{-1}$, $M_2 = q(1 + q)^{-1}$,
and $M_{\rm envelope} = 0$, where $q$ is the system mass ratio
given it Table 1.  But because the stars are being modeled on
a discrete computational mesh that does not conform precisely
to their shape, and because the acceleration of each fluid
element in the computational mesh is being determined by
finite-difference (rather than continuous differential)
representations of gradients in the pressure and gravitational
fields, as each system evolves hydrodynamically the surfaces of
the stars become less sharply defined and some spreading
of material inevitably occurs.  (In these benchmark evolutions,
this is evidenced for example by the very small, but finite
oscillations in the ``volumes'' occupied by the stars that
are displayed in Figs.\  \ref{fig:eb_vols} -
\ref{fig:ub_2_vols}.)  In practice, then, during each evolution
we determine whether material in each grid cell belongs to either
star or the ``envelope'' by comparing the binding energy of the
fluid in each cell to the average binding energy of the layer of
cells at the surface of each star.
In this context, we define the surface of each star to be the
layer of cells where the mass density falls below $10^{-5}$ in our normalized
units, which corresponds to the lowest density level attained in the initial
SCF models.
The mass of the envelope is dominated by material from the surface of the
stars even though there is a minimum ``vacuum'' density level of $1.0 \times 10^{-10}$
enforced by the code to maintain numerical stability.  The total mass of the
vacuum material is over a million times smaller than the mass of either
stellar component and does not impact the physics of these simulations.

\begin{figure}
    \plotone{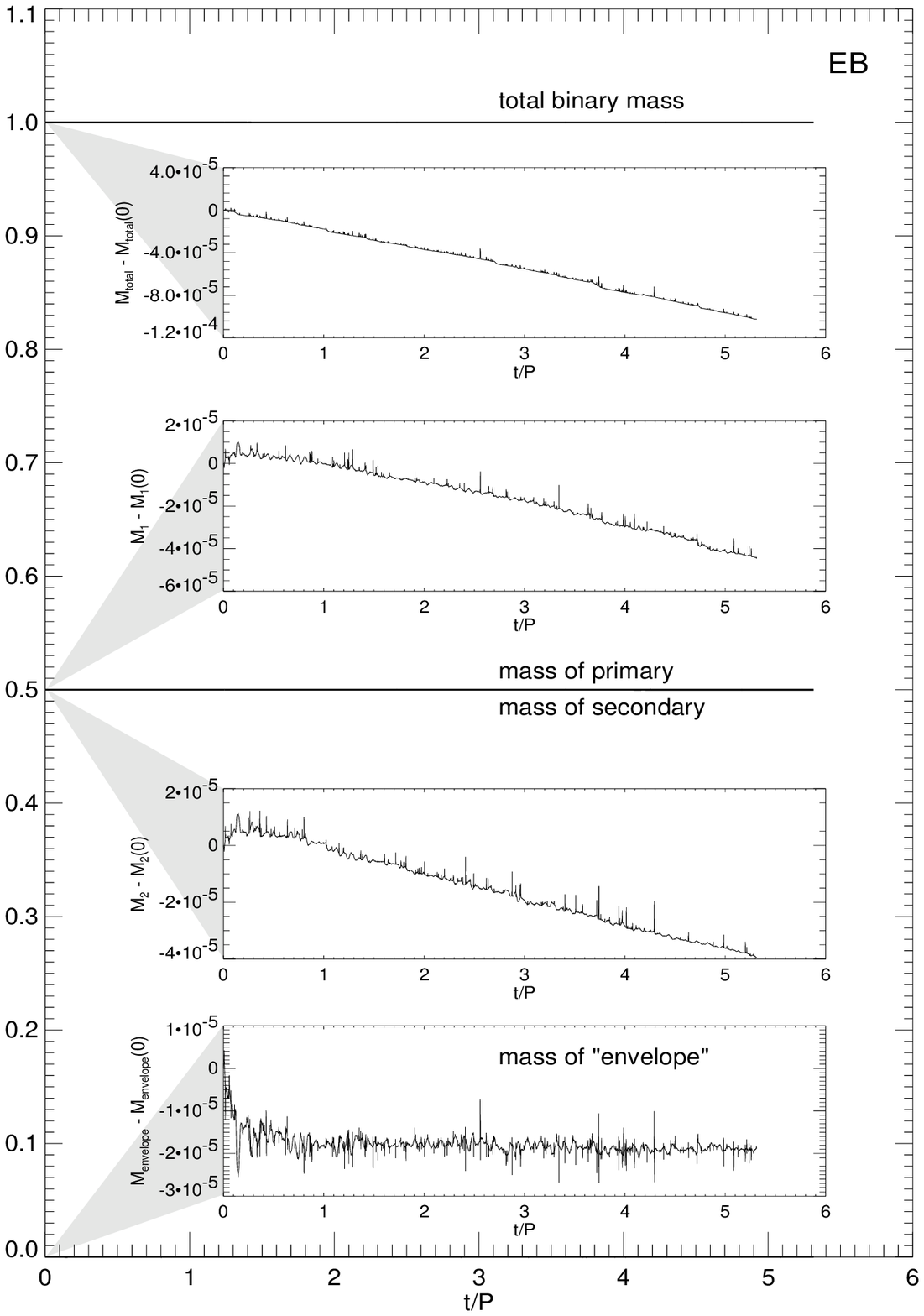}
    \figcaption[f12.eps]{Masses, normalized to the
    total system mass, plotted as a function of time, in units of the
    orbital period, for the EB simulation.  Top curve:  Total binary
    mass ($M$). Middle two curves:  Mass of the primary ($M_1$) and
    secondary ($M_2$) stars. Bottom curve (essentially at zero):  Mass
    of the ``envelope,'' as defined in \S 6.2. Inset plots show the
    difference between the indicated mass component and its initial
    value in units of the initial total mass.  \label{fig:eb_mass}}
\end{figure}

\begin{figure}
   \plotone{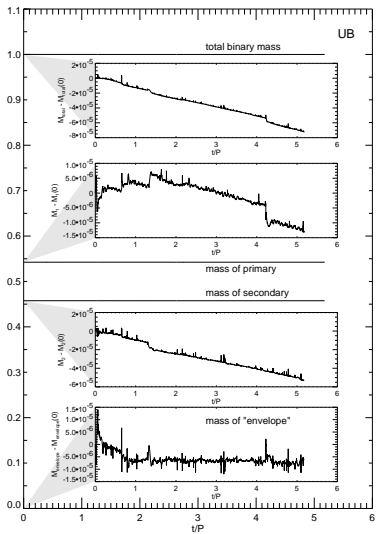}
   \figcaption[f13.eps]{Masses, normalized to the
   total system mass, plotted as a function of time, in units of the
   orbital period, for the UB simulation.  Top curve:  Total binary
   mass ($M$). Middle two curves:  Mass of the primary ($M_1$) and
   secondary ($M_2$) stars. Bottom curve (essentially at zero):  Mass
   of the ``envelope,'' as defined in \S 6.2. Inset plots show the
   difference between the indicated mass component and its initial 
   value in units of the initial total mass.  \label{fig:ub_mass}}
\end{figure}

Four curves are drawn (each at two quite different scales)
in Figs.\ \ref{fig:eb_mass} and \ref{fig:ub_mass} to document
how well mass is conserved in the EB and UB simulations,
respectively.   The masses have all been normalized to the
total system mass, so in Fig.\ \ref{fig:eb_mass} the mass
of both the primary ($M_1$) and the secondary ($M_2$) stars
is initially exactly $0.5$; the total binary mass is initially
exactly $1$; and the ``envelope'' mass is initially
exactly $0$.  In Fig.\ \ref{fig:ub_mass}, the total mass and
the ``envelope'' mass also are initially $1$ and $0$, respectively,
but the mass of the primary initially is $(1+q)^{-1} = 0.5424$
and the mass of the secondary initially is $q(1+q)^{-1} = 0.4576$.
Plotted on a normal, linear scale, all four of these curves are
perfectly flat in both figures.  This demonstrates that the
mass of both stars, as well as the aggregate system mass, is conserved
to very high precision throughout the EB and UB simulations.
Again, this is evidence that the initial models were in excellent
detailed force balance and the hydrodynamics code is evolving the
systems forward in time in a physically realistic manner.

Although mass is conserved to very high precision, it is
not absolutely constant throughout the evolutions.  In the four
insets of Figs.\ \ref{fig:eb_mass} and \ref{fig:ub_mass}, we
have magnified the vertical mass scale by roughly four orders of
magnitude in order to show that there is a very tiny, but
measurable, secular decrease in the total system mass and
in the mass of both stellar components over the course of the
simulations.  In each inset, we plot the relevant mass minus
its value in the initial state (time $t=0$), normalized to the
total system mass.  These inset plots show that the system
mass decreases by approximately one part in $10^4$ over five
orbits --- that is, about $0.002 \%$ per orbit --- with the mass
loss from each star accounting for roughly half this total.
In rows $2-4$ of Table \ref{tab:systems} we have recorded
for both evolutions
more precise values of the fractional mass that is lost, on
average, each orbit from the primary ($M_1$), the
secondary ($M_2$), and the system as a whole ($M$).
We have determined that this mass is very slowly lost as a result of
the development of a small, but nonzero flow of low-density material
off of the stars, through the envelope and, ultimately, off of the
grid. (After an initial drop, the
envelope mass remains approximately constant, suggesting that
this outward flow has settled into a nearly steady state.)
As is discussed more fully in \S7, below, this small but
detectable rate of mass loss from detached equilibrium
binaries imposes a straightforward
limit on the mass-transfer rates that we will be able to reliably
model in future simulations that involve dynamical mass transfer.

\subsection{Minimal Center-of-Mass Motion}

\begin{figure}
   \plotone{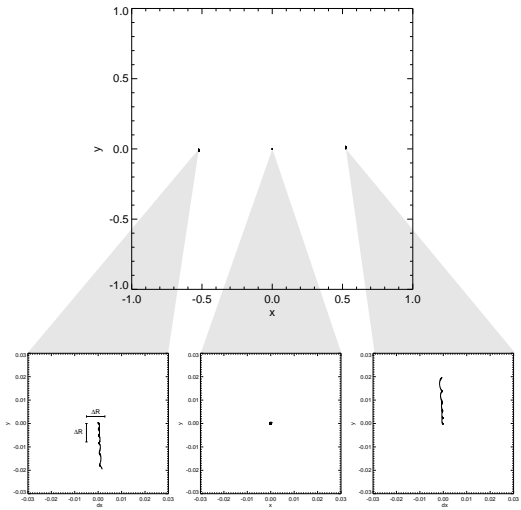}
   \figcaption[f14.eps]{From the EB simulation,
   equatorial-plane trajectories are plotted for the center of mass
   of the system and the centers of mass of both stellar components
   through just over $5$ orbits in the corotating frame of reference.
   Insets (from left to right) show magnified views of the
   trajectories for the secondary star, the system as a whole, and
   the primary star.  We have subtracted off the initial coordinates
   for the inset plots and have, for reference, indicated the size of
   one grid cell.  \label{fig:eb_com}}
\end{figure}

\begin{figure}
   \plotone{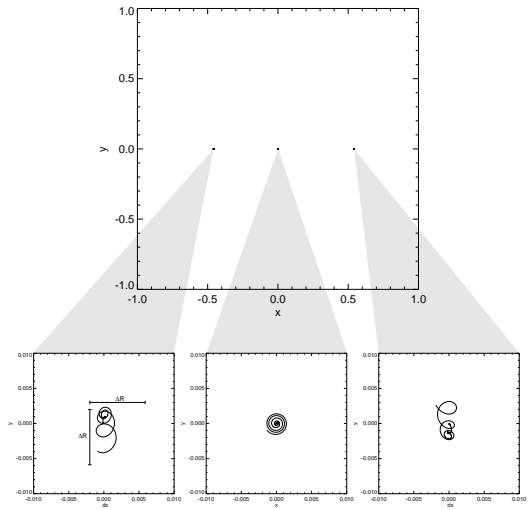}
   \figcaption[f15.eps]{From the UB simulation,
   equatorial-plane trajectories are plotted for the center of mass
   of the system and the centers of mass of both stellar components
   through just over $5$ orbits in the corotating frame of reference.
   Insets (from left to right) show magnified views of the 
   trajectories for the secondary star, the system as a whole, and
   the primary star.  We have subtracted off the initial coordinates
   for the inset plots and have, for reference, indicated the size of
   one grid cell.  \label{fig:ub_com}}
\end{figure}

During both simulations we also tracked as a function of time
the position of the center of mass of each binary component and
the position of the center of mass of the system as a whole.
The equatorial-plane trajectories of these three centers of
mass for the EB and UB evolutions are shown, respectively, in
Figs.\ \ref{fig:eb_com} and \ref{fig:ub_com},
as viewed from our computational reference frame
--- that is, from a frame rotating with the orbital angular velocity
of the system, as determined for the initial state by the SCF code.
In the uppermost plot of each figure, which has been drawn at
a scale ($-1 < x < +1$; $-1 < y < +1$) to include the entire mass
of the system, the three separate center of mass trajectories
appear to be small dots with little or no discernible structure.
(When plotted in the inertial reference frame on this scale the
trajectories of the two stars are indistinguishable from circles.)
This  illustrates that, even after five orbits, the
centers of mass of the two stars and of the system as a whole
essentially have not moved from their initial positions.
This provides additional strong confirmation that our SCF code
produces excellent initial states and that the hydrodynamical
equations are being integrated forward in time in a physically
realistic manner.

In the bottom three plots of Figs.\ \ref{fig:eb_com} and
\ref{fig:ub_com}, we have magnified a small region around
each of the center of mass trajectories ---
expanding the linear scale of the uppermost plot in each
figure by a approximately a factor of $15$ and $45$, respectively.
These magnified views reveal that, although it is very small,
there is measurable motion of the centers of mass in both
evolutions.  In the bottom, lefthand plot we also
have shown the size of our radial grid spacing,
$\Delta R = 7.87 \times 10^{-3}$.  This indicates the characteristic
size of our discretization and emphasizes how small the
motion of each center of mass is.  In the UB evolution, for example,
the motion of all three centers has been confined within a single
grid cell through five full orbits.  Furthermore, the smooth spiral
trajectory of the UB system center of mass (bottom, middle plot in
Fig. \ref{fig:ub_com}) has an understandable, physical origin.
As viewed in the inertial frame, this particular trajectory appears
as a straight line whose direction and magnitude is consistent
with the overall system velocity prescribed as initial conditions
from eq.\ (\ref{eq:com_vel}).  In the EB evolution, due to the
symmetry of the initial model, the drift of the system center of
mass is extremely small, remaining unnoticable even on the
magnified plot.  In the magnified plots,
the trajectory of the center of mass of each individual star shows
both a gradual drift in the $y$-direction, and a small oscillatory
motion in the $x$-direction.  The vertical drift is mostly an
indication that the binary's actual orbital frequency is slightly
different from the value (given by the SCF code) that we
used for the rotation frequency of the computational grid.
The oscillations in $x$ represent epicyclic motion and indicate that
the binary orbit is not precisely circular.  Since both the
drift and the epicyclic motion can be understood in physical terms,
their small amplitudes tell us more about the quality of the
initial model than about the limiting accuracy of our
finite-difference scheme.

Unlike in the single star case presented in \S
\ref{sec:equilibrium_test}, there is no evidence of a systematic
outwards force on either star in the UB or EB systems, despite the
fact that the systems have evolved for the equivalent of
approximately 90 dynamical times. It appears as though the
introduction of a rotating frame of reference and the associated
centrifugal potential and Coriolis force has provided a feedback
mechanism that acts to limit the systematic imbalance discussed
previously.

\subsection{Binary Separation}

\begin{figure}
   \plotone{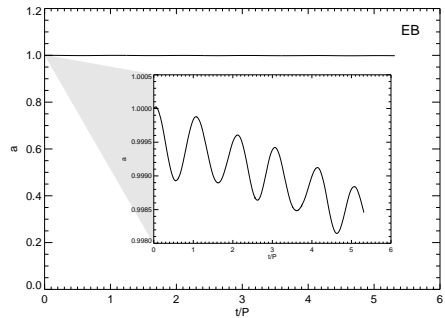}
   \figcaption[f16.eps]{The orbital separation, normalized to its initial value, as a
   function of orbital time for the EB system.  \label{fig:eb_a}}
\end{figure}

\begin{figure}
   \plotone{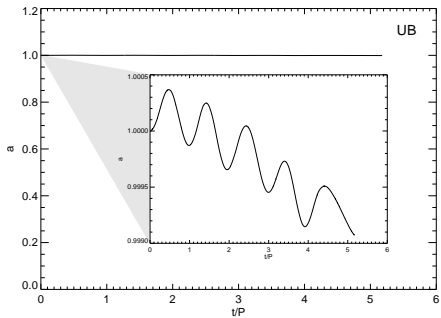}
   \figcaption[f17.eps]{The orbital separation, normalized to its initial value, as
   a function of orbital time for the UB system. \label{fig:ub_a}}
\end{figure}

A plot of the binary separation $a$ as a function of time, as
shown in Figs.\ \ref{fig:eb_a} and \ref{fig:ub_a} for the EB and
UB evolutions, respectively, provides another way to assess the
global behavior of these systems.  Here, the separation $a$ is
defined as the distance between the centers of mass of the two
stars.  Notice that, on a linear
scale that extends from $0$ to $1$ (in units normalized to the
each system's initial separation), the plot of $a(t)$ is
indistinguishable from a perfectly horizontal line.  This illustrates
that, to a very high degree of accuracy, these benchmark simulations
of detached binary systems produce stable, circular orbits.

Again, though, if we examine these plots in finer detail, we see
that both evolutions exhibit a very small but quantifiable
departure from perfect circular orbital motion.  For example, in
the insets to Figs.\ \ref{fig:eb_a} and \ref{fig:ub_a}, we have
reploted $a(t)$ with the vertical scale magnified by roughly a
factor of $400$.  These insets show that in both evolutions there
is a very slow, secular decrease in the orbital separation and, in
addition, $a(t)$ displays low-amplitude oscillations having a
period approximately equal to one orbital period.  The
oscillations in $a$ arise from the same epicyclic motion that was
seen in the plots (Figs.\ \ref{fig:eb_com} and \ref{fig:ub_com})
of the center of mass motion of the individual stars, but the
amplitude of this motion is easier to measure here. In units of
the initial orbital separation, the EB system has an epicyclic
amplitude $(\Delta a/a)_{\rm epicyclic} \approx 5 \times 10^{-4}$;
the UB system exhibits an epicyclic amplitude about half this
size. The slow, secular decay of the orbits occurs at a rate
$({\Delta a}/{a})_{\rm secular} \approx 2.9 \times 10^{-4}$ per
orbit in the EB system, and at a rate ${\Delta a}/{a} \approx  1.9
\times 10^{-4}$ per orbit in the UB system. These orbital decay
rates and epicyclic amplitudes have been recorded in the fifth and
sixth rows of Table \ref{tab:systems}.

\subsection{Angular Momentum Conservation}

\begin{figure}
   \plotone{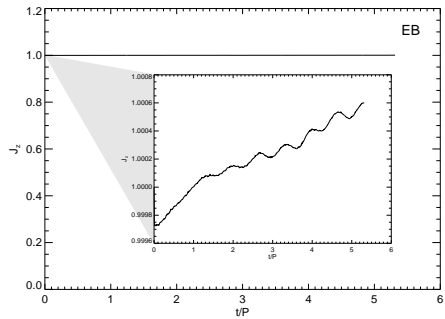}
   \figcaption[f18.eps]{The z component of total angular momentum, normalized to
   its initial value, as a function of time for the EB system. \label{fig:eb_jz}}
\end{figure}

\begin{figure}
   \plotone{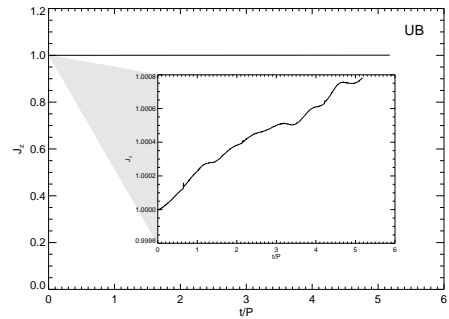}
   \figcaption[f19.eps]{The z component of total angular momentum, normalized to
   its initial value, as a function of time for the UB system.  \label{fig:ub_jz}}
\end{figure}

Finally, in Figs.\ \ref{fig:eb_jz} and \ref{fig:ub_jz} we show the
behavior as a function of time of the $z$-component of each
system's total angular momentum.  As was true with our plots of
the orbital separation, on a linear scale that extends from $0$ to
$1$ (in units normalized to the each system's initial total
angular momentum), the plot of $J_z(t)$ is indistinguishable from
a perfectly horizontal line.  This illustrates that these
benchmark simulations globally conserve angular momentum to a very
high degree of accuracy.  When we magnify the vertical scale by
approximately a factor of $1000$, as has been done to produce the
insets to Figs.\ \ref{fig:eb_jz} and \ref{fig:ub_jz}, we see that
angular momentum is not, in fact, perfectly conserved. Evidently
both systems gain angular momentum at a very slow rate; in the EB
simulation, ${\Delta J_z}/{J_z} \approx 1.1 \times 10^{-4}$ per
orbit, and in the UB simulation ${\Delta J_z}/{J_z} \approx 1.5
\times 10^{-4}$ per orbit. These rates have been recorded in the
seventh row of Table \ref{tab:systems} and will be referred to
again in \S7, below, when we summarize the limiting accuracy with
which we expect to be able to model physical mass-transfer events
using our simulation tools.

\subsection{Overview}

We should emphasize that the hydrodynamics code as described in \S
4 and utilized in these benchmark simulations has evolved through
many stages from the version of the code that was used several
years ago by \citet{new97} to investigate the equal-mass, binary
merger problem.  A number of improvements were made in the code in
order to bring it to its present level of performance.  Figure 20
is presented here in an effort to illustrate how certain key
modifications in the code affected its performance.  Each row of
frames in this figure shows results from an evolution of the same
unequal-mass (UB) binary system that was used in our benchmark
simulation, but as produced by four separate versions of the code.
The curves drawn in the four frames on the left-hand side of Fig.\
20 show the same type of information as has been presented in Fig.\
10 for the benchmark UB evolution:  Four separate volumes for the
secondary star (solid curves) and its Roche volume (dashed curve)
are plotted as a function of time, in units of the orbital period.
The four frames on the right-hand side of Fig.\ 20 show the same
type of information as has been presented in the top-most frame of
Fig.\ 15:  Center-of-mass trajectories of the two stars and of the
system as a whole, as viewed from the rotating frame of reference.
(The bottom-most frames are taken from the benchmark UB simulation
and therefore are drawn directly from Figs.\ 10 and 15.)

\begin{figure}
   \plotone{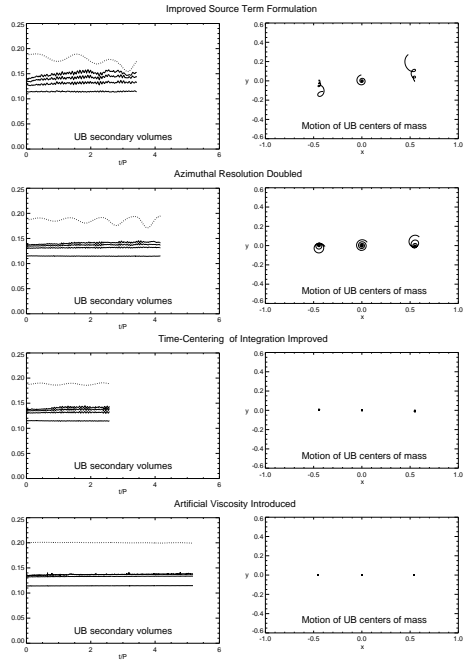}
   \figcaption[f20.eps]{Each of the four pairs of plots shown here
   has been derived from the UB simulation as modeled with one of
   four separate versions of our hydrodynamic code. In each pair, the
   plot on the left is directly analogous to Fig. 10, showing as a
   function of time the Roche volume (dashed curve) and four interior
   volumes (solid curves) for the secondary star; the plot on the 
   right is directly analogous to the unmagnified plot in Fig. 15,
   showing the trajectories of the center of mass of the secondary
   (left), system as a whole (center), and primary (right).  The
   pairs of plots are shown chronologically from the top to the
   bottom, with significant improvements in the code being made
   between each recorded UB simulation.  See \S6.6 for a description
   of these various code improvements. \label{fig:code_devel}}
\end{figure}

The results shown in the top-most frames of Fig.\ 20 come from an
early version of the code in which we replaced the gradient of the
pressure with the gradient of the enthalpy.  This ensured that the
initial structure of each star, as determined by the SCF code, was
in good force balance after being introduced into the hydrocode.
However, as the two figures from this evolution illustrate, we
still observed a slow expansion of the secondary star; the orbit
itself developed a significant epicyclic amplitude; and after
about three orbits, the Roche lobe was encroaching on the surface
of the secondary. The second row of frames comes from a simulation
in which the number of azimuthal zones was doubled --- from 128 to
256 zones over the full $2\pi$ radians.  This modification
improved somewhat the mean motion of the centers of mass (although
it did not significantly reduce the amplitude of the epicyclic
motion).  Most significantly, however, doubling the angular grid
size improved the resolution and, hence, the determination of
force balance in each star. As a result, expansion of the
secondary star was noticeably reduced.  The third row of frames
shows that motion of the centers of mass was drastically reduced
when we modified our algorithm to make the integration scheme more
properly time-centered.  This change did not noticeably reduce the
rate of expansion of the secondary, but it did significantly
reduce the amplitude of oscillations in the Roche lobe volume.
Finally, by introducing artificial viscosity into the equations of
motion in order to mediate the weak shocks at the surface of the
stars (which also involved a re-centering of all momentum densities
to the cell locations specified in Table 3), the entire structure
of both stars became much more robust.
In particular, as the left-hand frame of the last row shows (see
also Fig. 10), this code modification completely eliminated the
short timescale wiggles in the volumes of the secondary; overall
expansion of the secondary also was reduced to an imperceptible
level.  Simultaneously, for the first time, we ascribed a small
nonzero velocity to the initial state as given by eq. (60).
This change further reduced the motion of the centers
of mass --- to the level illustrated by Fig.\ 15.

It is reasonable to ask whether the three principal spurious
effects that remain in our benchmark simulations --- the slow
decay of the orbits, the slow gain of angular momentum, and the
slow loss of mass from the stars --- are at least mutually
consistent on physical grounds. For centrally condensed binaries,
a point mass approximation (the Roche approximation) is usually
sufficient to discuss the orbital evolution. This approximation
predicts a simple relationship between the changes of mass,
angular momentum and binary separation, namely
\begin{equation}
   \frac{\Delta J_{\rm com}}{J_{\rm com}} = \frac{1}{2}
      \frac{\Delta a}{a} +
      \frac{\Delta M_1}{M_1}  +  \frac{\Delta M_2}{M_2} -
      \frac{1}{2} \frac{\Delta M}{M} \, ,
\end{equation}
where $J_{\rm com}= M_1 M_2 \sqrt{G a/M}$ is the total
center--of--mass angular momentum in the point mass approximation.
Numerical values for each of the terms in this expression can be
obtained from the data shown in Table 6.  In particular, we see
that in both benchmark simulations the three mass terms
approximately cancel each other out.  But while the magnitude of
$(\Delta a/a)_{\rm secular}$ is roughly the same as the magnitude
of $\Delta J_z/J_z$, their signs are different.  That is, the
angular momentum of the system is slowly increasing while the
binary separation is slowly decreasing.  This is clearly
inconsistent with the expectations of the simple Roche model. A
more accurate expression for the total angular momentum of the
binary would be
\begin{equation}
  J_{\rm bin} = J_{\rm com} + I_1 \Omega_1 + I_2 \Omega_2\, ,
\end{equation}
where $I_i$ and $\Omega_i$ are the moments of inertia and inertial
frame angular velocities of the binary components, assuming they
all rotate around the same $z$-axis. Even if one takes into
account the contributions of spin angular momenta, the changes
observed remain inconsistent and must therefore be attributed to
spurious numerical effects at a level of $10^{-4}$ per orbit
arising from the inevitable error terms present in our
finite-difference representation of the fluid equations.  What we
have attempted to do here is quantitatively document the magnitude
of these numerical effects in the highest practical resolution
possible at the present day for simulations of detached binaries
where the character of the ideal solution is well understood
beforehand. Furthermore, we can not accurately predict the
evolution of mass transferring binaries where the mass transfer
rate, ${\Delta M}/{M}$ per orbit is $\lesssim {\rm few} \times
10^{-5}$ using simulations at the resolution presented here. There
are however a wide variety of systems (the initial mass transfer
event in an Algol progenitor, or the onset of common envelope
evolution in the progenitors of many types of binaries, or the
formation of Type Ia supernovae for example) that are expected to
exceed our threshold resolution limit for mass transfer.  At a
sufficiently high mass-transfer rate, the mass transfer itself
will drive the evolution of the orbital parameters and Roche
geometry at a rate higher than the numerical limits demonstrated
here.

\section{Conclusions}

In this paper we have presented a practical SCF algorithm for
constructing self-consistent polytropic binaries with unequal
masses that satisfy the condition of hydrostatic equilibrium to a
high degree of accuracy.  This three-dimensional SCF algorithm is
based largely on the technique first described by
\citet{hachisu86}, but to our knowledge this is the first time the
technique has been used to generate unequal mass binary systems as
input for a hydrodynamical simulation.  Our two benchmark
simulations (described in \S 6) clearly indicate that this SCF
algorithm can provide superb initial states for investigations
into the dynamical stability of close binary systems.  We
emphasize that, in addition to generating models of close binary
systems that are detached --- like the EB and UB systems
constructed for our two benchmark simulations --- as shown above
in Figs.\ \ref{fig:scf_side} and \ref{fig:scf_top} this technique
also can be used to generate close binary systems that are
semi-detached or, in the limit of identical components, in
contact.

We have also detailed our gravitational hydrodynamics code and
presented results from key tests of the stability and accuracy of
the hydrodynamics algorithm, the solution of Poisson's equation,
and the coupled solution required to evolve an isolated, spherical
polytrope that is placed off-axis in a cylindrical
grid.  From these test cases it is apparent that a number of
subtle numerical issues arise when a highly nonaxisymmetric body
is evolved via an explicit integration of finite-difference equations
on a cylindrical computational grid.  It also appears however, that
these effects can be made manageably small by increasing
the resolution used to treat the system of interest.

We have evolved two detached binary
systems (one with a mass ratio $q = 1$, the other with a mass ratio
$q=0.8436$) through more than five orbits in order to benchmark
in detail the capabilities of our simulation tools.
Even though the individual stellar components generated by our
SCF code are significantly rotationally flattened (due to the synchronous
rotation of the initial states) and tidally distorted (by their close
binary companion), these benchmark simulations show that the stars are
in almost perfect hydrostatic equilibrium.  Throughout each binary
evolution, our hydrodynamics code conserves mass and angular momentum
to a very high degree of precision; as viewed from a frame of reference
that is rotating with the initial orbital frequency of the binary,
the centers of mass of the two stellar components and of the system
as a whole remain virtually stationary; and a plot of the binary
separation as a function of time shows that the stellar orbits are
almost indistinguishable from circles.  This gives us considerable
confidence that these numerical tools can be used to examine the
stability of close binary systems against both tidal and mass-transfer
instabilities, and to begin to accurately model mass
transfer in semi-detached systems.

As has been summarized in Table \ref{tab:systems}, from our
benchmark evolutions we have been able to determine in
quantitative terms the level of accuracy with which our
hydrodynamical code conserves mass, conserves angular momentum,
and is able to represent and maintain a circular binary orbit.
Mass is conserved to roughly $0.002\%$ per orbit; angular momentum
is conserved to a level of $0.01$ - $0.02\%$ per orbit; the binary
separation remains constant to a few parts in $10^4$ per orbit;
and each orbit exhibits an epicyclic amplitude (measured relative
to the orbital separation) of $.02$-$.05\%$.

We are unaware of any other group that is attempting to study the
onset of mass-transfer instabilities in unequal-mass binaries with
a gravitational hydrodynamics code, like ours, that fully resolves
both stellar components.  Hence, there are no published numbers
against which to compare ours for the UB evolution.  However, we
can fairly compare the results of our EB evolution against the
recent study published by SWC of equal-mass close binary systems.
Their Fig.\ 10 illustrates that, after following one stable binary
system through approximately $6$ orbits (we assume, based on the SWC
discussion, that $P= 1.7 - 2\  {\rm ms}$), they have been able to
conserve angular momentum to a level of about $0.2\%$ per orbit.
And their Fig.\ 14 shows four stable orbits with epicyclic
amplitudes (measured relative to the binary separation) $\sim 0.3
- 1\%$.  We conclude that, at least in these two respects, our
simulations improve on the SWC models by roughly one order of
magnitude.  SWC do not comment on their level of mass
conservation; and, because of the visible epicyclic motions in
their Fig.\ 14, it is difficult to ascertain to what degree the
binary separation either decreases or increases with time over the
course of their evolutions. In both of our evolutions, however,
the center-of-mass motion of our stars appears to be significantly
less than the center-of-mass motion depicted for SWC's preferred
integration scheme in the top-left panel of their Fig.\ 7.

The small, but measurable changes in mass, angular momentum, and
binary separation documented here in Table \ref{tab:systems} set
limits on the types of mass-transfer events that we will be able
to model with confidence using our present simulation tools.  For
example, if we were to try to model an instability that leads to a
flow through the binary's L1 Lagrange point with a mass-transfer
rate lower than one part in $10^6$ per orbit, the physical
exchange of material between the binary components would be
swamped by the unphysical process that is causing our stars to
lose mass to the ``envelope'' at a rate of one to two parts in
$10^5$ per orbit. If the depth of contact between the Roche lobe
and the surface of the donor star is not sufficient, epicyclic
motion in the orbit will tend to shut off the mass-transfer during
part of each orbit. Also, a drift in the center of mass of the
system can impose a limit on the length of time that the binary
can be evolved before the motion of the binary through the grid
becomes problematic.  We will have to contend with all of these
issues as we move to the next level of our investigation and
introduce a semi-detached system from our SCF code into our
hydrodynamical code. We expect nevertheless to find a wide range
of interesting binary systems whose dynamical evolution can be
simulated in a fully self-consistent fashion through a reasonably
large number of orbits using the tools that have been described in
this paper.

As we have documented in Table \ref{tab:systems}, the calculation
of one orbit takes about 33 wall-clock hours when utilizing 64
processors of the Cray T3E 600, and using 16 processors of the
newer IBM SP-3, the calculation of one orbit takes about 51 hours.
The computational workload of a mass-transfer simulation is
therefore within
the reach of current, state of the art, parallel computers given
the linear scaling of our gravitational hydrodynamics code with
the number of processors even at a resolution greater than
presented here.  We note that the amount of work performed can be
reduced significantly if need be by, for example, freezing the
gravitational potential until the mass distribution has changed
significantly as done by \citet{stone92}.  The solution of
Poisson's equation represents about a quarter of the total
execution time.

We have been able to estimate the mass transfer rate required to
bring the simulation  above the level of the noise observed in
our benchmark simulations. We find that this value is
$\sim \mathrm{few} \times10^{-5}$ of the donor's initial mass over an
orbital period.  As discussed in \S \ref{sec:stability}, the mass
transfer rate should scale as a high power of the degree of
over-contact (as the cube for an $n = 3 / 2$ polytrope);
furthermore, for a case where $q > 1$, that is, the donor is
initially the more massive star, the degree of over-contact will
naturally increase with time as the star expands and its Roche
lobe shrinks.  Provided that such a binary system can begin
mass-transfer, the amplitude of the mass transfer rate should
inevitably reach values higher than indicated above.  
Since motion of the center of mass of the binary system
has been confined to a region well within a single computational
grid cell even after $5$ orbits, we are confident that future
evolutions can be followed with confidence through at least
$20 - 30$ orbits, given sufficient computing resources.
As discussed
in the introduction of this paper, we understand that the
mass transfer rates considered here are much larger than those
found in what are considered typical examples of interacting
binaries.  The methods presented here are not applicable to the
stable mass transfer observed in cataclysmic variables or other
long lived systems, but should serve very well to investigate
stages of evolution of their progenitors and
transient events such as the onset of dynamical mass transfer and
its stability.

\acknowledgments
This work has been performed with support from the National Science Foundation
through grants AST-9720771, AST-9528424, AST-9987344, and DGE-9355007 and from the
National Aeronautics and Space Administration through the Astrophysics
Theory Program grant NAG5 8497.  This research has been supported,
in part, by grants of high-performance computing time at the National
Partnership for Advanced Computing Infrastructure (NPACI) facilities at
the San Diego Supercomputer Center and by Louisiana State University's
High Performance Computing facilities.  We would also like to acknowledge
the many useful comments made by the referee that led to a significant improvement in
the contents of this paper.


\begin{thebibliography}{}
\bibitem[Batten(1989)]{batten89} Batten, A.\ H.\ 1989, Proc.\ IAU Coll.\ Nr.\ 107, \textit{Algols} (Dordrecht:Kluwer Academic Publishers)

\bibitem[Bhattacharya(1995)]{bhattacharya95} Bhattacharya, D.\ 1995, in \textit{X-Ray Binaries}, ed.\
W.\ H.\ G.\ Lewin, J.\ van Paradijs, \& E.\ P.\ J.\ van den Heuvel (Cambridge: Cambridge University Press), 233

\bibitem[Bisikalo et al(2000)]{bisikalo00} Bisikalo, D.\ V., Harmanec, P., Boyarchuk, A.\ A., Kuznetov, O.\ A., \& Hadrava, P.\ 2000, \aap, 353, 1009

\bibitem[Black \& Bodenheimer(1975)]{black75} Black, D.\ C., \& Bodenheimer, P.\ 1975, \apj, 199, 619

\bibitem[Blondin, Richards \& Malinowski(1995)]{blondin95} Blondin, J.\ M., Richards, M.\ T., \& Malinowski, M.\ L.\ 1995, \apj, 445, 939

\bibitem[Boroson et al(2001)]{boroson01} Boroson, B., Kallman, T., Blondin, J.\ M., \& Owen, M.\ P.\ 2001, \apj, 550, 919

\bibitem[Bowers \& Wilson(1991)]{bowers91} Bowers, R.\ L., \& Wilson, J.\ R.\ 1991, \textit{Numerical Modeling in Applied Physics and Astrophysics} (Boston:Jones and Bartlett)

\bibitem[Cazes \& Tohline(2000)]{cazes00} Cazes, J.\ E., \& Tohline, J.\ E.\ 2000, \apj, 532, 1051

\bibitem[Chandrasekhar(1939)]{chandrasekhar39} Chandrasekhar, S.\ 1939, \textit{An Introduction to the Study of Stellar Structure} (Chicago: University of Chicago Press)

\bibitem[Cohl, Sun \& Tohline(1997)]{cohl97} Cohl, H.\ S., Sun, X.-H., \& Tohline, J.\ E.\ 1997, in Proceedings of the 8th
SIAM conference on Parallel Processing for Scientific Computing, ed. M.\ Heath, V.\ Torczon, G.\ Asffalk, P.\ E.\ Bj\o rstad, A.\ H.\ Karp, C.\ H.\ Koebel, V.\ Kumar, R.\ F.\ Lucas, L.\ T.\ Watson, \& D.\ E.\ Womble (Philadelphia: SIAM)

\bibitem[Cohl \& Tohline(1999)]{cohl99} Cohl, H.\ S., \& Tohline, J.\ E.\ 1999, \apj, 527, 86

\bibitem[Frank, King \& Raine(2001)]{frank01} Frank, J., King, A.\ R., \& Raine D.\ J.\ 2001, \textit{Accretion Power
in Astrophysics}, (3rd ed:Cambridge: Cambridge University Press), in press

\bibitem[Hachisu(1986)]{hachisu86} Hachisu, I.\ 1986, \apjs, 62, 461

\bibitem[Hachisu, Eriguchi \& Nomoto(1986)]{hachisu86b} Hachisu, I., Eriguchi, Y., \& Nomoto, K.\ 1986, \apj, 311, 214

\bibitem[Hawley, Wlison \& Smarr(1984)]{hawley84} Hawley, J.\ F., Wilson, J.\ R., \& Smarr, L.\ L.\ 1984, \apjs, 55, 211

\bibitem[Iben \& Livio(1993)]{iben93} Iben, I., Jr., \& Livio, M.\ 1993, PASP, 105, 1373

\bibitem[Iben \& Tutukov(1984)]{iben84} Iben, I., Jr., \& Tutukov, A.\ V.\ 1984, \apjs, 54, 335

\bibitem[Iben(1990)]{iben90} Iben, I., Jr.\ 1990, \apj, 353, 215

\bibitem[Janka et al(1999)]{janka99} Janka, H.-T., Eberl, T.\, Ruffert, M.\,  \& Fryer, C.\ L.\ 1999 \apjl, 527, L39

\bibitem[King et al(1997)]{king97} King, A.\ R.\, Frank, J.\, Kolb, U.\, \& Ritter, H.\ 1997, ApJ, 482, 919

\bibitem[Lai, Rasio \& Shapiro(1994)]{lai94} Lai, D.\, Rasio, F.\ A.\, \& Shapiro, S.\ L.\ 1994, \apj, 423, 344

\bibitem[Lewin, van Paradijs \& van den Heuvel(1995)]{lewin95} Lewin, W.\ H.\ G.\, van Paradijs, J.\, \& van den Heuvel, E.\ P.\ J.\ 1995 \textit{X-Ray Binaries}, (Cambridge: Cambridge University Press)

\bibitem[Lindblom, Tohline \& Vallisneri(2001)]{lindblom01} Lindblom, L.\, Tohline, J.\ E.\, \& Vallisneri, M.\ 2001, \prl, 86, 1152L

\bibitem[Lufkin \& Hawley(1993)]{lufkin93} Lufkin, E.\ A.\, \& Hawley, J.\ F.\ 1993, \apjs, 88, 569

\bibitem[Marietta, Burrows \& Fryxell(2000)]{marieta00} Marietta, E.\, Burrows, A.\, \& Fryxell, B.\ 2000, \apjs, 128, 615

\bibitem[M\'esz\'aros(2001)]{meszaros01} M\'esz\'aros, P.\ 2001, Science, 291, 79

\bibitem[Nelson \& Eggleton(2001)]{nelson01} Nelson, C.\ A.\ \& Eggleton, P.\ P.\ 2001, ApJ, 552, 664

\bibitem[New \& Tohline(1997)]{new97} New, K.\ C.\ B.\, \& Tohline, J.\ E.\ 1997, \apj, 490, 311

\bibitem[New, Centrella \& Tohline(2000)]{new00} New, K.\ C.\ B.\, Centrella, J.\ M.\, \& Tohline, J.\ E.\ 2000, \prd, 620, 13

\bibitem[Norman \& Winkler(1983)]{norman83} Norman, M.\ L., \& Winkler, K.-H.\ 1983, in \textit{Astrophysical
Radiation Hydrodynamics},  ed. K.-H.\ Winkler and M.\ L.\ Norman (Dordrecht:Reidel), 449

\bibitem[Paczy\'{n}ski(1971)]{paczynski71} Paczy\'{n}ski, B. 1971, \araa, 9, 183

\bibitem[Paczy\'{n}ski(1986)]{paczynski86} Paczy\'{n}ski, B. 1986, \apjl  308, L43

\bibitem[Paczy\'{n}ski and Sienkiewicz(1972)]{paczynski72} Paczy\'{n}ski, B.\, \& Sienkiewicz, R.\  1972, Acta Astronomica, 22, 74

\bibitem[Peaceman \& Rachford(1955)]{peaceman55}Peaceman, D.\, \& Rachford, H.\ 1955, J. Soc. Indust. Appl. Math., 3, 28

\bibitem[Ruffert et al(1997)]{ruffert97} Ruffert, M., Janka, H.-T., Takahashi, K.,  \& Schaefer, G.\ 1997, \aap, 319, 122

\bibitem[Sandquist et al(1998)]{sandquist98} Sandquist, E.\ L., Taam, R.\ E., Xingming, C., Bodenheimer, P., \&  Burkert, A.\ 1998, \apj, 500, 909

\bibitem[Schwarztrauber \& Sweet(1975)]{schwarztrauber75} Schwarztrauber, P., \&  Sweet R.\ 1975, in
Efficient FORTRAN Subroutines for the Solution of Elliptic Partial Differential Equations,
NCAR Technical Note - TN/IA-109

\bibitem[Schwarztrauber et al(2001)]{schwarztrauber00} Schwarztrauber, P., Sweet, R., \& Adams, J.\
     \url{http://www.scd.ucar.edu/css/software/fishpack}

\bibitem[Stone \& Norman(1992)]{stone92} Stone, J.\ M., \& Norman, M.\ L.\ 1992, \apjs, 80, 753

\bibitem[Sod(1978)]{sod78} Sod, G.\ 1978, J.\ Comput.\ Phys., 27, 1

\bibitem[Swesty et al(2000)]{swesty2000} Swesty, F.\ D., Wang, E.\ Y.\ M., \& Calder, A.\ C.\ 2000, \apj, 541 937 (SWC)

\bibitem[Trimble(1983)]{trimble83} Trimble, V.\ 1983, \nat, 303, 137

\bibitem[van Leer(1979)]{vanleer79} van Leer, B.\ 1979, J.\ Comput.\ Phys., 32, 101

\bibitem[Verbunt \& van den Heuvel(1995)]{verbunt95} Verbunt, F., \& van den Heuvel, E.\ P.\ J.\ 1995, in \textit{X-ray Binaries}, ed.\
W.\ H.\ G.\ Lewin, J.\ van Paradijs, \& E.\ P.\ J.\ van den Heuvel (Cambridge: Cambridge University Press), 457

\bibitem[Vesper, Honeycutt \& Hunt(2001)]{vesper01} Vesper, D., Honeycutt, K., \& Hunt, T.\ 2001, AJ, 121, 2723

\bibitem[Warner(1995)]{warner95} Warner, B.\ 1995, \textit{Cataclysmic Variable Stars}, (Cambrdige: Cambridge University Press)

\end{thebibliography}
\end{document}